\documentclass[11pt,a4paper]{article}
\usepackage{silence} 
\usepackage{jcappub}
\usepackage[english]{babel}
\usepackage[utf8]{inputenc}
\usepackage[OT2,T1]{fontenc}
\usepackage{graphicx}
\usepackage{orcidlink} 
\usepackage{gensymb}
\usepackage{amsmath,amssymb,amsfonts,amsthm}
\usepackage{bm,bbm}
\usepackage{tikz,stackrel}
\usepackage{pstricks}
\usepackage{pifont} 
\usepackage{empheq}
\newcommand\nn{\nonumber\\}
\newcommand\non{\nonumber}
\newcommand{\bc}{\begin{center}}
	\newcommand{\ec}{\end{center}}
\newcommand{\be}{\begin{equation}}
	\newcommand{\ee}{\end{equation}}
\newcommand{\ba}{\begin{eqnarray}}
	\newcommand{\ea}{\end{eqnarray}}
\def\bs{\begin{subequations}}
	\def\es{\end{subequations}}
\newcommand{\ben}{\begin{equation*}}
	\newcommand{\een}{\end{equation*}}
\newcommand{\ban}{\begin{eqnarray*}}
	\newcommand{\ean}{\end{eqnarray*}}
\renewcommand{\leq}{\leqslant}
\renewcommand{\geq}{\geqslant}
\def\a{\alpha}
\def\b{\beta}

\def\Om{\Omega}
\def\om{\omega}

\def\s{\sigma}

\def\cN{\mathcal{N}}

\def\cP{\mathcal{P}}

\def\cT{\mathcal{T}}

\def\p{\partial}

\newcommand{\Eq}[1]{(\ref{#1})}
\newcommand{\Eqq}[1]{eq.~(\ref{#1})}
\newcommand{\Eqqs}[1]{eqs.~(\ref{#1})}

\def\cob{\color{blue}}

\newcommand{\au}[2]{#1.~#2}
\newcommand{\book}[5]{\emph{#1}, #2, #3, #4 (#5)}

\newcommand{\oarX}[1]{\href{http://arxiv.org/abs/#1}{{\ttfamily\cob arXiv:#1}}}
\newcommand{\arX}[1]{\href{http://arxiv.org/abs/#1}{{\ttfamily\cob arXiv:#1}}}
\newcommand{\doin}[6]{\href{http://dx.doi.org/#1}{{\cob {\it #2} {\bf #3 #4} (#6) #5}}}
\newcommand{\doinn}[5]{\href{http://dx.doi.org/#1}{{\cob {\it #2} {\bf #3} (#5) #4}}}
\newcommand{\doij}[5]{\href{http://dx.doi.org/#1}{{\cob {\it #2} {\bf #3} (#5) #4}}}

\newcommand{\ndoinn}[5]{\href{#1}{{\cob {\it #2} {\bf #3} (#5) #4}}}

\newcommand{\tia}[1]{\textit{#1},}

\def\Mpl{M_{\rm Pl}}

\def\rmd{d}

\newcounter{listcounter}

\allowdisplaybreaks

\def\laq{~\raise 0.4ex\hbox{$<$}\kern -0.8em\lower 0.62ex\hbox{$\sim$}~}
\def\gaq{~\raise 0.4ex\hbox{$>$}\kern -0.7em\lower 0.62ex\hbox{$\sim$}~}

\def\beq{\begin{equation}}
	\def\eeq{\end{equation}}
\def\bea{\begin{eqnarray}}
	\def\eea{\end{eqnarray}}

\def \ra {\rightarrow}

\def \Mp {\Mpl}

\def \da {\delta}
\def \b {\beta}
\def \a {\alpha}
\def \ap {\alpha^{\prime}}

\def \sg {\sigma}

\def \da {\delta}
\def \ep {\epsilon}
\def \r {\rho}
\def \om {\omega}
\def \Om {\Omega}

\begin{document}
	
	\renewcommand{\thefootnote}{\fnsymbol{footnote}}
	
	\title{Gravitational-wave background in bouncing models from semi-classical, quantum and string gravity}
	
	\author[a]{Ido Ben-Dayan\,\orcidlink{0000-0002-6292-3981},}
	\emailAdd{idobd@ariel.ac.il}
	\affiliation[a]{Physics Department, Ariel University, Ariel 40700, Israel}
	
	\author[b,*]{Gianluca Calcagni\,\orcidlink{0000-0003-2631-4588}\note{Corresponding author.},}
	\emailAdd{g.calcagni@csic.es}
	\affiliation[b]{Instituto de Estructura de la Materia, CSIC, Serrano 121, 28006 Madrid, Spain}
	
	\author[c,d]{Maurizio Gasperini\,\orcidlink{0000-0001-9117-8303},}
	\emailAdd{gasperini@ba.infn.it}
	\affiliation[c]{Istituto Nazionale di Fisica Nucleare, Sezione di Bari, Via G.\ Amendola 173, 70126 Bari, Italy}
	\affiliation[d]{Dipartimento di Fisica, Universit\`a di Bari, 
		Via G.\ Amendola 173, 70126 Bari, Italy}
	
	\author[e]{Anupam Mazumdar\,\orcidlink{0000-0002-0967-8964},}
	\emailAdd{anupam.mazumdar@rug.nl}
	\affiliation[e]{Van Swinderen Institute, University of Groningen, 9747 AG Groningen, The Netherlands}
	
	\author[c,d]{Eliseo Pavone\,\orcidlink{0000-0002-3022-4545},}
	\emailAdd{eliseo.pavone@ba.infn.it}
	
	\author[a,f]{Udaykrishna Thattarampilly\,\orcidlink{0000-0001-9170-4641},}
	\emailAdd{uday7adat@gmail.com}
	\affiliation[f]{Astrophysics Research Center, The Open University of Israel, Raanana, 43107, Israel}
	
	\author[a]{Amresh Verma\,\orcidlink{0009-0003-0372-2359}}
	\emailAdd{amreshv@ariel.ac.il}
	
	\abstract{We study the primordial spectra and the gravitational-wave background (GWB) of three models of semi-classical, quantum or string gravity where the big bang is replaced by a bounce and the primordial tensor spectrum is blue: ekpyrotic universe with fast-rolling Galileons, string-gas cosmology with Atick--Witten conjecture and pre-big-bang cosmology. We find that the ekpyrotic scenario with Galileons does not produce a GWB amplitude detectable by present or third-generation interferometers, while the Atick--Witten-based string-gas model is ruled out in its present form for violating the big-bang-nucleosynthesis bound, contrary to the original string-gas scenario. In contrast, the GWB of the pre-big-bang scenario falls within the sensitivity window of both LISA and Einstein Telescope, where it takes the form of a single or a broken power law depending on the choice of parameters. The latter will be tightly constrained by both detectors. 
	}
	
	
	\maketitle
	\renewcommand{\thefootnote}{\arabic{footnote}}
	
	
	\section{Introduction}\label{sec1}
	
	Gravitational waves (GWs) are one of the newest and most exciting windows into the workings of gravity. Thanks to the observation of GWs from binary systems by the LIGO-Virgo-KAGRA (LVK) network of ground-based interferometers \cite{KAGRA:2021vkt,LIGOScientific:2021sio}, and of a gravitational-wave background (GWB) $\Omega_{\textsc{gw}}(f)\sim f^{2/3}$
	of possibly astrophysical origin in the International Pulsar Timing Array (IPTA) \cite{NANOGrav:2023gor,NANOGrav:2023hvm,EPTA:2023fyk,Reardon:2023gzh,Xu:2023wog,InternationalPulsarTimingArray:2023mzf,Ben-Dayan:2023lwd}, we have acquired a better understanding of the physics of neutron stars and black-hole binaries, as well as of the behaviour of the gravitational force near these compact objects and of its propagation across cosmological distances. Both LVK and third-generation instruments such as the Laser Interferometer Space Antenna (LISA) \cite{Barausse:2020rsu,LISACosmologyWorkingGroup:2022jok,LISA:2022kgy,Colpi:2024xhw}, Einstein Telescope (ET) \cite{Maggiore:2019uih,Branchesi:2023mws} and DECIGO \cite{Seto:2001qf,Kawamura:2011zz,Kawamura:2020pcg} will be able to further probe Einstein's theory and its extensions to modified gravity and quantum gravity \cite{Mirshekari:2011yq,Canizares:2012is,Ellis:2016rrr,Yunes:2016jcc,Arzano:2016twc,Calcagni:2019kzo,Cardoso:2019rvt,Belgacem:2019pkk,Calcagni:2019ngc,Auclair:2019wcv,Belgacem:2020pdz,Calcagni:2020tvw,Addazi:2021xuf,LISACosmologyWorkingGroup:2022wjo,Calcagni:2022tuz,Baker:2022eiz,Calcagni:2023vxg}.
	
	The detection of a relic primordial GWB \cite{Christensen:2018iqi,Renzini:2022alw,vanRemortel:2022fkb} would be a momentous opportunity to look into the early universe and the gravitational interaction in extreme curvature or energy regimes. At the level of fundamental physics it is not easy to construct robust cosmological models embedded in realistic scenarios of modified or quantum gravity, and even less so to obtain one such model predicting an observable signal without invoking an \emph{ad hoc} matter field dynamics. With a first, rapid scan of the literature we may find five candidates attempting to fulfil these characteristics, heterogeneous in terms of robustness and predictive power \cite{Calcagni:2020tvw}: nonlocal Starobinsky inflation \cite{Koshelev:2016xqb,Koshelev:2017tvv}, Brandenberger--Ho non-commutative inflation \cite{Brandenberger:2002nq,Calcagni:2013lya}, the S-brane ekpyrotic universe \cite{Brandenberger:2020tcr,Brandenberger:2020eyf,Brandenberger:2020wha}, string-gas cosmology \cite{Brandenberger:2006xi,Brandenberger:2011et,Brandenberger:2012um,Brandenberger:2014faa,Bernardo:2020nol,Bernardo:2020bpa} and multi-fractional inflation \cite{Calcagni:2016ofu}. However, only the last four are possibly able to produce a detectable signal and only crossing the DECIGO sensitivity curve in the most optimistic cases. A sixth model of quantum gravity, nonlocal and non-minimally coupled with radiation, appeared afterwards with similar characteristics \cite{Calcagni:2022tuz}. Recently, however, a more detailed exploration of the landscape of quantum and string  cosmologies  \cite{LISACosmologyWorkingGroup:2022jok} singled out three more scenarios with a signal potentially reaching LISA and the Einstein Telescope:
	\begin{itemize}
		\item A bouncing model with a slow ekpyrotic contraction phase sustained by fast-rolling Galileons with a non-canonical kinetic term and where perturbations are sourced by a $U(1)$ gauge field \cite{Cai:2012va,Ben-Dayan:2016iks,Ben-Dayan:2018ksd,Artymowski:2020pci,Ben-Dayan:2023rlj}. This particle content is the main difference with respect to the S-brane ekpyrotic scenario of \cite{Brandenberger:2020tcr,Brandenberger:2020eyf,Brandenberger:2020wha}. While the original ``old'' ekpyrotic scenario \cite{Khoury:2001wf} was inspired by string-theory concepts, the current model is based on effective quantum field theory, and is not necessarily tied to a specific quantum-gravity theory.
		\item A bouncing model where the contracting phase is dominated by a string gas behaving like a stiff fluid and evolving according to Einstein's gravity \cite{Biswas:2014kva}. The main difference with respect to the string-gas cosmology of \cite{Brandenberger:2006xi,Brandenberger:2011et,Brandenberger:2012um,Brandenberger:2014faa,Bernardo:2020nol,Bernardo:2020bpa} is that, while the latter model is based on the behaviour of closed-string modes below the Hagedorn temperature, in \cite{Biswas:2014kva} the string thermodynamics was studied above the Hagedorn temperature. This implies that the free energy of the strings grows with the temperature more slowly than for ordinary radiation.
		\item A pre-big-bang model evolving from the  string perturbative vacuum, proposed long ago \cite{Gasperini:1992em,Gasperini:2002bn,
			Gasperini:2007vw,Gasperini:2016gre} on the grounds of the string cosmological equations, which enjoy T-duality \cite{3} 
		and may be characterized by a non-singular bounce thanks to all-orders (higher-curvature) $\ap$ corrections \cite{7,8}.
	\end{itemize}
	The first two models are phenomenological because they are not fully embedded in any high-energy quantum theory of gravity. The third one is based on the modified gravitational dynamics uniquely fixed, in principle, by the string unification of all fundamental interactions, at all energy scales including their quantum limit; however, it also contains phenomenological aspects concerning the (presently unknown) details of the dilaton dynamics in the strong coupling regime. In any case, all three models above are among the very few in quantum gravity possibly able to produce a GWB, arising from the vacuum fluctuations of the metric tensor, with high enough amplitude in the frequency range of third-generation detectors. They have in common an initial phase of growing curvature scale (described by a contracting kinematics with the metric of the Einstein frame), preceding the final decelerated expansion and passing through a non-singular bounce of spacetime curvature.
	The presence of accelerated contraction ($\dot a <0$, $\ddot a <0$, $\dot H<0$) or of super-inflationary expansion ($\dot a >0$, $\ddot a >0$, $\dot H>0$) in different metric frames is responsible for a strong blue tilt in the primordial tensor spectrum (i.e., growing with the frequency). 
	
	The first two models have been mainly explored at the level of the primordial tensor and scalar spectra, while for the third an approximate GWB profile is known (first computed in \cite{4,5,6,19} and recently discussed in \cite{Gasperini:2016gre}). The GWB of none of them, however, has been studied systematically so far, and the question of whether their signal can reach LISA, ET or DECIGO remains open. Also, one may wonder whether the common characteristic of having a contracting phase would produce a unique type of signal. It is the purpose of this paper to give an answer to these questions and to complement the analysis of \cite{Calcagni:2020tvw} done for other models of quantum gravity with a blue primordial tensor spectrum. We find that:
	\begin{itemize}
		\item The GWB amplitude of the ekpyrotic model with Galileons is highly suppressed and unobservable, contrary to the S-brane ekpyrotic scenario of \cite{Brandenberger:2020tcr,Brandenberger:2020eyf,Brandenberger:2020wha}.
		\item The string-gas cosmological model following the Atick--Witten conjecture is ruled out because its signal is too high and violates current bounds, contrary to string-gas models not adopting this conjecture
		\cite{Brandenberger:2011et,Brandenberger:2012um,Bernardo:2020nol,Bernardo:2020bpa}.
		\item Within the theoretically allowed parameter space, the GWB of the pre-big-bang model reaches the LISA and ET observational window while respecting present bounds. In all plots, we use the latest ET sensitivity curve for a single 10 km triangular interferometer with a signal-to-noise ratio 1 and a one-year observation run \cite{Branchesi:2023mws,ETsensitivity}.
	\end{itemize}
	To the best of our knowledge, the pre-big-bang model and those studied in \cite{Calcagni:2020tvw} are the only ones motivated by quantum gravity and generating a detectable GWB directly from the primordial tensor sector. Recently, another bouncing model was proposed where curvature perturbations evolving through a bounce can trigger the formation of primordial black holes and also induce a GWB signal with high amplitude crossing also the LISA and ET windows \cite{Papanikolaou:2024fzf}. However, we do not consider scalar-induced GWBs here.
	
	The paper is organized as follows. Basic expressions connecting the primordial tensor spectrum and the GWB are recalled in section \ref{sec11}. The primordial spectra and the GWB of the three models above are studied, respectively, in sections \ref{sec2}, \ref{sec3} and \ref{sec4}. Conclusions are in section \ref{sec5}. Technical material is relegated to appendices.
	
	
	\subsection{Basic formul\ae}\label{seccon}
	\label{sec11}
	
	Primordial GWs can be described by a small set of quantities and observables, independently of the underlying model. We denote the primordial tensor and scalar spectra as, respectively, $\cP_{\rm t}(k)$ and $\cP_{\rm s}(k)$, where $k$ is the comoving wave-number. From here, one calculates the tensor-to-scalar ratio at any given pivot scale $k=k_*$,
	\be\label{r}
	r\coloneqq \frac{\cP_{\rm t}}{\cP_{\rm s}}\,,
	\ee
	as well as the tensor and scalar spectral indices
	\be
	n_{\rm t} \coloneqq \frac{\rmd\ln\cP_{\rm t}(k)}{\rmd\ln k}\,,\qquad
	n_{\rm s}-1 \coloneqq \frac{\rmd\ln\cP_{\rm s}(k)}{\rmd\ln k}\label{nsnt}\,.
	\ee
	The current estimate for the scalar amplitude is $\cP_{\rm s}(k_*)\approx 2.1\times10^{-9}$ at the cosmic microwave background (CMB) scale $k_*=0.05\,{\rm Mpc}^{-1}$ \cite{Akrami:2018odb}, while $n_{\rm s}=0.9649 \pm 0.0042$ at 68\% confidence level at the same pivot scale, assuming $\rmd n_{\rm s}/\rmd\ln k=0$ \cite{Planck:2018vyg,ParticleDataGroup:2022pth}. The upper bound on \Eq{r} is $r(k_*)<0.036$ at the same scale at $95\%$ confidence level \cite{BICEP:2021xfz}.
	
	In general, the amplitude of the GWB is defined as
	\beq\label{41}
	\Om_{\textsc{gw}} (k,\tau_0) = \frac{1}{\r_{\rm crit}(\tau_0)} \frac{\rmd \r_k (\tau_0)}{\rmd \ln k}\,,
	\eeq
	where $\tau_0$ is the present value of the conformal time $\tau$, defined in terms of the cosmic time $t$ as $\tau\coloneqq\int\rmd t/a(t)$, where $a(t)$ is the scale factor, $\r_{\rm crit} = 3 \Mp^2 H^2$ is the critical density, $\Mpl^2= (8 \pi G)^{-1}$ is the reduced Planck mass and $\r_k (\tau_0)$ is the energy density of the $k$-th Fourier mode of tensor perturbations amplified by the given model of the early universe and evaluated at the present time $\tau_0$. For GWs generated by tensor perturbations, the GWB spectral shape can be recast as
	\be\label{Omgw}
	\Om_\textsc{gw}(f) =\frac{k^2}{12a_0^2H_0^2}\cP_{\rm t}(k)\,\cT^2(k, \tau_0)\Big|_{k=2\pi f}\,,
	\ee
	where $f=k/(2\pi)$ is the GW frequency measured in Hz, $a_0$ is the scale factor today ($a_0=1$), $H_0$ is the value of the Hubble parameter today and $\cT(k,\tau_0)$ is the transfer function encoding how the primordial spectrum evolved after horizon crossing until today \cite{Turner:1993vb,Watanabe:2006qe,Kuroyanagi:2008ye}. The expressions below are valid for any model where observable perturbations have re-entered the horizon in the past, either because they were originally generated inside the horizon and where later expelled out (e.g., by inflation), or because they were generated outside the horizon in the first place. In the case of instantaneous reheating and a quick transition to a radiation-dominated phase, which applies to all the models discussed below, one has \cite{Nakayama:2008wy} (see \cite{Figueroa:2019paj} for an alternative way to express the transfer function)
	\be\label{Tk}
	\cT^2 (k, \tau_0) =  
	\Omega_{{\rm m}0}^2 \left[ \frac{g_\ast ( T_{\rm in})}{g_{\ast 0}} \right]
	\left[ \frac{g_{\ast s0}}{g_{\ast s} (T_{\rm in})} \right]^{4/3}
	\left[ \frac{ 3 j_1 (k\tau_0)}{k \tau_0} \right]^2 \cT_{\rm eq}^2 (k) \,,
	\ee 
	where $j_1$ is the first spherical Bessel function, the fitting function $g_\ast (T_{\rm in})$ is \cite{Kuroyanagi:2014nba}
	\ba
	g_\ast [T_{\rm in}(k)]&=&g_{\ast 0} 
	\frac{A_1+\tanh \left\{-2.5\log_{10}\left[\frac{k/(2\pi)}{2.5\times 10^{-12} ~{\rm Hz}}\right]\right\}}{A_1+1}\\
	&&\times
	\frac{A_2+\tanh \left\{-2.0\log_{10}\left[\frac{k/(2\pi)}{6.0\times 10^{-9} ~{\rm Hz}} \right] \right\}}{A_2+1}\,,\label{gast}
	\ea
	with $g_{\ast 0}=3.36$, $A_1=(-1-10.75 /g_{\ast 0})/(-1+10.75/g_{\ast 0})$, $A_2=(-1-g_{\rm max}/10.75) /(-1+g_{\rm max}/10.75)$, $g_{\rm max}=106.75$ for a Standard-Model particle content and the function $g_{\ast s}(T_{\rm in})$ is the same as \Eq{gast} upon replacing $g_{\ast 0}\to g_{\ast s0}=3.91$. The fitting function at the end of \Eq{Tk} is \cite{Turner:1993vb}
	\be
	\cT_{\rm eq}^2(k) = 1 + 1.57 \frac{k}{k_{\rm eq}} + 3.42 \left(\frac{k}{k_{\rm eq}}\right)^2\,,\qquad k_{\rm eq} = 7.1 \times 10^{-2}\, \Omega_{\rm m} h^2~{\rm Mpc}^{-1},
	\ee
	$k_{\rm eq}$ being the comoving wave-number at radiation-matter equality, so that  $f_{\rm eq}\approx 9.9\times 10^{-17}$ $(\Omega_{\rm m} h^2/0.143) \,{\rm Hz}$, where $\Om_{\rm m}=0.3153$ is the matter-energy density and $h$ is the dimensionless Hubble parameter, which we take at its CMB-scale value $h= 0.6736$ \cite{Akrami:2018odb}.
	
	
	\section{Sourced bounce with fast-rolling Galileons}\label{sec2}
	
	In the article \cite{Ben-Dayan:2023rlj}, a non-singular bounce model was proposed as an alternative model for structure formation in the early Universe. Non-singular bounce models often predict a deeply blue scalar spectrum contrary to CMB observations\footnote{A notable exception is the  model of \cite{Wands:1998yp} where a matter bounce produces nearly scale-invariant scalar and tensor spectra.}. While some ingredients are similar to the ekpyrotic scenario, such as a slow contraction with an equation of state $w \gg 1$, others are very different, as the observed density and tensor perturbations on CMB scales are generated by sourced fluctuations, rather than vacuum ones, amounting to a different paradigm. A specific model was devised using a gauge field resulting in gravitational waves with a specific chirality \cite{Ben-Dayan:2016iks}. The difference in the spectral index was shown to be remedied by the inclusion of that gauge field \cite{Artymowski:2020pci,Ben-Dayan:2018ksd}; the scalar spectrum matches current observations and the model predicts a tensor-to-scalar ratio with values below the upper bound \cite{Ben-Dayan:2023rlj}
	\be
	r\lesssim 10^{-2}\,.
	\ee
	The model involves Galileons with a non-canonical kinetic term and a coupling with a $U(1)$ gauge field. The action is
	\begin{equation}
		S = \Mpl^2\int \rmd^4x \sqrt{|g|} \left[\frac{R}{2}-K(\phi,X)+G(\phi,X)\Box \phi-I^2(\phi) \left( \frac{1}{4}F^{\mu\nu}F_{\mu \nu} -\frac{\delta}{4}\tilde{F}^{\mu \nu}F_{\mu\nu}\right) \right], \label{eq:vecaction}
	\end{equation}
	where $\Mpl$ is the reduced Planck mass, $\phi$ is the Galileon field, $A_{\mu}$ is the $U(1)$ gauge vector,  $F_{\mu\nu} = \partial_\mu A_\nu - \partial_\nu A_\mu$, $\tilde{F}^{\mu\nu} = \frac{1}{2}\epsilon^{\mu\nu\rho\sigma}F_{\rho\sigma}$ and $\delta>0$ is a coupling constant. The functions $K$ and $G$ in \Eq{eq:vecaction} are
	\ba
	&& K(\phi,X) = [1-g(\phi)] X+ \beta X^2-V(\phi)\,, \label{eq:kin}\\
	&& G(\phi,X) = \gamma X\,,\\
	&&g(\phi) = \frac{2g_0}{e^{-\sqrt{\frac{2}{p}}\phi}+e^{b_g\sqrt{\frac{2}{p}}\phi}} \, , \;\;\;\;\;V(\phi) = -\frac{2V_0}{e^{-\sqrt{\frac{2}{q}}\phi}+e^{b_V\sqrt{\frac{2}{q}}\phi}}\,,
	\ea
	where $X=-\frac{1}{2}\partial_{\mu}\phi \partial^{\mu} \phi$ and $\beta$, $\gamma$ and $g_0$ are parameters. The background dynamics of the Universe is determined by a single scalar field $\phi$ with a non-trivial kinetic term typical of Galileons. This model, which is different from others involving Galileons (e.g., \cite{Choudhury:2023hfm,Choudhury:2024one}), gives rise to a non-singular bounce and the potential $V$ is chosen in such a way as to obtain ekpyrotic contraction away from the bounce.  The Universe starts at $\phi \ll -1$, far away from the bounce, with a slow ekpyrotic contraction. As $\phi$ accelerates towards $\phi$ = 0, the value of $g$ increases. Since $g(0)> 1 $  (which we require), at some point in time, $g$ exceeds the critical value $g = 1$ and the sign of the kinetic term $X$ in \eqref{eq:kin}
	becomes negative, giving rise to a phase of ghost condensation coinciding with the bouncing phase. This is the region in field space where the null energy condition is violated,\footnote{Although the null energy condition is violated, we confirm that the average null energy condition is not violated.} which in turn triggers the bounce at $\phi=0$. 
	The Universe continues to roll to positive larger values of the field $\phi>\phi_{B+}$, after which the Lagrangian regains a canonical form and the Universe enters an era of kinetic energy domination. $I$, the coupling of the scalar field to the gauge field, is 
	\begin{equation}
		I(\phi) = \frac{1}{1+e^{-a_1 n (\phi-\phi_{B-})}}
		\label{eq:I}\,,
	\end{equation}
	where $a_1 =1/\sqrt{2q}$ and $\phi_{B-}$ is value of field $\phi$ at the beginning of the bounce. $I(\phi)$ is defined such that during the regime of ekpyrosis, i.e., for large and negative $\phi$, 
	\begin{equation}
		I(\phi) \simeq e^{a_1 n \phi}.
	\end{equation}
	The values $n=2$ or $n=-1$ lead to scale-invariant sourced perturbations \cite{Artymowski:2020pci}. 
	Introducing gauge fields can source second-order inhomogeneities that dominate over the first-order perturbations (characterized by a blue spectrum) at CMB scales. This leads to sourced and unsourced perturbations that are linearly independent since the equation for perturbations is linear and creation/annihilation operators of sourced and unsourced fluctuations are uncorrelated. The total power spectrum is of the form 
	\begin{equation}
		\mathcal{P}^{\rm tot} = \mathcal{P}^v + \mathcal{P}^s
	\end{equation}
	where $\cP^v$ is the unsourced or vacuum spectrum and $\mathcal{P}^s$ is the sourced spectrum. The vacuum scalar ($\cP_{\rm s}^v$) and tensor ($\cP_{\rm t}^v$) spectra as well as the sourced scalar ($\cP_{\rm s}^s$) and tensor ($\cP_{\rm t}^s$) spectra are \cite{Ben-Dayan:2023rlj}
	\begin{equation}
		\begin{split}
			\cP^v_{\rm t}(k) &= \left(\frac{k}{H_{B^-}} \right)^{\frac{6(1+w_{\rm ekp})}{1+3w_{\rm ekp}}} \frac{\gamma_{E}^2 H_{B^-}^2}{2 \pi^2 \Mpl^2}\,, \\
			\cP^v_{\rm s}(k) &= \left(\frac{k}{H_{B^-}} \right)^{\frac{6(1+w_{ekp})}{1+3w_{ekp}}}  \frac{\gamma_{E}^2 H_{B^-}^2}{48 \pi^2 \Mpl^2} F_{\zeta}^2\,, \\
			\cP^s_{\rm t}(k) &= 2.8 \frac{9 e^{4\pi \xi}}{8 \pi q^4 \xi^6} \left(\frac{H_{B-}}{\Mpl}\right)^4 \left(\frac{k}{H_{B-}}\right)^{n_{\rm t}}\,, \\
			\cP^s_{\rm s}(k)  &=  2.8 \frac{225 e^{4\pi \xi}}{32 \pi q^4 \xi^6} \left(\frac{H_{B-}}{\Mpl}\right)^4 \left(\frac{k}{H_{B-}}\right)^{n_{\rm s}-1} F_{\zeta}^2\,,\\
			\label{e:finspec1}
		\end{split}
	\end{equation}
	where $\gamma_{E}$ is the Euler--Mascheroni constant, $q$ is related to the equation of state of the ekpyrotic phase by $w_{\rm ekp} = -1+ \frac{2}{3q}$, $\xi$ is related to the strength of gauge coupling by $\xi= n \delta$ and $H_{B-}$ is the Hubble parameter at the end of the bounce determined by the scale of the bounce. $n_{\rm s}$ and $n_{\rm t}$ are the spectral indices of scalar and tensor perturbations, respectively. By choosing $n=-2.01$, it is possible to obtain a scalar spectral index $n_{\rm s}=0.96$, a red tilted scalar power spectrum in agreement with observations. The tensor-to-scalar ratio $r$ is determined by the factor $F_{\zeta}$, governing amplification across the bounce. An approximate expression for $F_{\zeta}$ is given by
	\begin{equation}
		F_{\zeta}= \cosh\left[\int_{\tau_{B}^-}^{\tau_{B}^+} \rmd\tau'\,\omega_S(\tau')\right], 
		\label{eq:amp}
	\end{equation}
	where $\tau_{B}^-$ and $\tau_{B}^+$ denote the beginning and end of the bounce and $\omega_S^2 = {z''}/{z}$, where $z$ during bounce phase is approximately $z=a^2 3 \beta/(\gamma_E^2\dot{\phi}^2)$. It is not easy to write a closed-form expression for $F_{\zeta}$, except for the case where we assume $\dot{\phi}$ to be Gaussian. In \cite{Ben-Dayan:2023rlj}, the authors evaluated $F_{\zeta}$ semi-analytically. This enhancement is driven by the $\omega_S^2 = {z''}/{z}$ term that dominates over $c_s^2k^2$. 
	The bounce is short enough so that the evolution of modes remains under perturbative control, even for the high-$k$ modes. For high-$k$ modes it was checked numerically, while for other modes it was also verified analytically. All modes are amplified by the same finite amount. That the growth of perturbations does not lead to divergences as long as the bounce duration is short can be verified numerically.
	
	Here lies a main qualitative difference between the different spectra. The scalar spectra are amplified across the bounce while the tensor ones are unchanged. Since the sourced spectrum is significantly larger compared to the vacuum spectrum, $\mathcal{P}^{\rm tot} \simeq \mathcal{P}^s$ and
	\begin{equation}
		r \simeq \frac{\cP_{\rm t}^s}{\cP_{\rm s}^s} = \frac{4}{25} \left( \frac{1}{F_{\zeta}} \right)^2.
	\end{equation}
	The sourced tensor spectrum has the same red-tilted spectral index as the scalar spectrum and is unobservable by current and third-generation interferometers \footnote{It should be noted that this sourced tensor spectrum is observable in CMB frequencies. It is distinguishable from the inflationary predictions by the fact that the signal should have a definite chirality, unlike the inflationary predictions that result in similar contribution to the spectrum from both chiralities.}. However, the vacuum perturbations are characterized by a blue spectrum, and although tensor vacuum perturbations are small compared to the sourced spectrum at CMB scales, the blue spectrum may be observable at higher frequencies. 
	
	After the bounce, kinetic domination takes place. The gauge field in our model can be either a putative gauge field, or the actual $U(1)$ of electromagnetism. The different options will imply different reheating epochs and may modify the predicted tensor spectrum.
	We now examine the possibility of the gauge field sourcing perturbations as electromagnetic radiation. It is possible to calculate the scale factor at which the Universe ends a reheating phase with this assumption. The Universe becomes radiation dominated when the energy density of the scalar field responsible for the bounce is smaller than the energy density of the gauge field. The energy density of the scalar field $\phi$ at the end of the bounce phase is approximately 
	\begin{equation}
		\rho_{\phi}(\tau_{B+}) \simeq \frac{1}{2} \dot{\phi}^2(\tau_{B+})[1-g_0+ 3 \beta \dot{\phi}^2(\tau_{B+})] \simeq \frac{g_0-1}{3\beta} e^{-2\frac{\tau_{B+}}{T}} \left(1-e^{-2\frac{\tau_{B+}}{T}}\right)\,,
	\end{equation}
	where $\tau_{B+}$ and $\tau_{B-}$ are the conformal time at the beginning and at the end of the bounce phase, respectively, and $T$ is a quarter of the duration of bounce phase. During the phase of kinetic domination, the energy density decays as $a^{-6}$. Thus, the energy density of the scalar field during this phase is
	\begin{equation}
		\rho_{\phi} = \frac{g_0-1}{3\beta} e^{-2\frac{\tau_{B+}}{T}} \left(1-e^{-2\frac{\tau_{B+}}{T}}\right) \left(\frac{a_B}{a}\right)^6.
	\end{equation}
	The gauge field energy density attains the maximum
	\begin{equation}
		\rho_{A,\tau_{B-}} = D(n) \frac{ e^{2\pi \xi} }{\xi^3 \tau_{B-}^4}\,,\qquad D(n)=\frac{1}{4\pi^2} \frac{(n-1)^2 \Gamma(2n-1)}{2^{2n+1}(n-2)\pi}\,,
	\end{equation}
	at the bounce, where the expression for $D(n)$ holds for $n$ close to 2. After the bounce, the energy density decays as $a^{-4}$ and the gauge field energy density is 
	\begin{equation}
		\rho_{A,\tau} = D(n)  \frac{ e^{2\pi \xi} }{\xi^3 \tau_{B-}^4} \left(\frac{a_B}{a}\right)^4\,.
	\end{equation}
	From the expressions for energy densities, one can calculate the scale factor corresponding to which the Universe ends the reheating phase and find that reheating will last for only a few e-folds. 
	
\begin{figure}[ht]
		\centering
		\includegraphics[width=12cm]{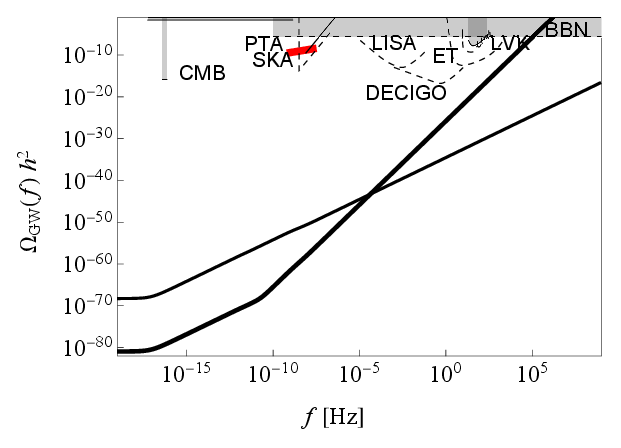}\\
		\includegraphics[width=12cm]{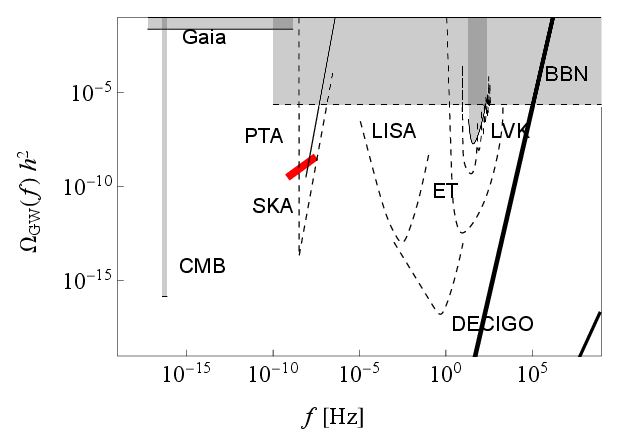}
		\caption{GWB generated by the primordial vacuum tensor spectrum of the ekpyrotic model with fast-rolling Galileons for the instantaneous reheating scenario with $H_{B+}=H_{\rm r}= 10^{-5}\,\Mpl$ (solid thin curve) and the prolonged reheating scenario with $H_{B+}=10^{-5}\,\Mpl$ and $H_{\rm r}=10^{-43}\,\Mpl$
			(solid thick curve), compared with the sensitivity curves (dashed) of LVK, SKA, LISA, ET and DECIGO. Also shown are the CMB and BBN exclusion (where the CMB upper bound is given thickness to improve visibility in the plot) and, as a thick red segment, the signal detected by PTAs. The bottom plot captures the high-amplitude part of the GWB.\label{fig1}}
	\end{figure}

	In the derivation of primordial spectra in \eqref{e:finspec1}, we have assumed the scale factor at the bounce to be one. However, the transfer function is derived with normalization $a_0=1$. For consistency, we reparametrize $a_0$ as
	\begin{equation}
		a_0 = a_B \frac{a_{\rm r}}{a_{B+}} \frac{a_{B+}}{a_B} \frac{a_0}{a_{\rm r}}
		\label{eq:a0}
	\end{equation}
	where $a_{\rm r}$ is the scale factor at the onset of radiation domination. The reheating phase for this model is dominated by the kinetic term from $t_{B+}$ to the beginning of radiation domination, implying 
	\begin{equation}
		\frac{a_{\rm r}}{a_{B+}} = \left( \frac{t_{\rm r}}{t_{B+}} \right)^{\frac13} \simeq \left( \frac{H_{B+}}{H_{\rm r}} \right)^{\frac13}.
		\label{eq:a1}
	\end{equation}
	From \eqref{eq:a0} and \eqref{eq:a1}, we obtain
	\begin{eqnarray}
		\frac{a_0}{a_B} = \frac{34 \times 10^{13}}{87 \, \text{GeV}} H_{B+}^{\frac13} H_{\rm r}^{\frac16} \Mpl^{\frac12}\,.
	\end{eqnarray}
	Instantaneous reheating implies $H_{\rm r}\simeq H_{B+}$. In our case, the value of $H_{B+}$ is determined by the theory of sourced perturbations and is dependent on the parameters of the theory. In our discussion, we set parameters for sourced perturbations to values optimal for obtaining a reasonable tensor-to-scalar ratio and a red spectral tilt. This value is around $H_{B+}\simeq 10^{-5} \Mpl$. We notice that, for the case where the gauge field responsible for sourcing gravitational waves behaves as radiation, the reheating phase is short, $\mathcal{N}\sim 3\!-\!4$ e-folds, and the spectrum of GWs is similar to that of the case where reheating is instantaneous. For models with instant reheating, the transfer function is described in \eqref{Tk} and present-day GWB is given by \eqref{Omgw}. We present our results in Fig.\ \ref{fig1}, where we used vacuum tensor spectrum from \eqref{e:finspec1} assuming $w_{\rm ekp} \gg 1$ which gives\footnote{This expression applies to most ekpyrotic models such as the old scenario \cite{Khoury:2001wf}, the new scenario with instabilities \cite{Lehners:2007ac,Buchbinder:2007ad,Buchbinder:2007at,Buchbinder:2007tw} and the new scenario without instabilities \cite{Qiu:2013eoa,Li:2013hga,Fertig:2013kwa,Ijjas:2014fja} but not to the S-brane scenario \cite{Brandenberger:2002nq,Calcagni:2013lya,Brandenberger:2020tcr}.}
	\be\label{ekpyTT}
	\cP_{\rm t}^v(k) \simeq  \frac{\gamma_{E}^2 k^2}{2 \pi^2 \Mpl^2}\,.
	\ee
	
	If, instead, we assume that the gauge field responsible for sourced perturbations is not radiation and is something more exotic such as $U(1)$ axions or dark radiation, then it is possible to have an extended reheating period where the Universe is dominated by a kinetic term. We have also examined this possibility. The transfer function in the presence of a reheating phase is slightly modified and is well known \cite{Nakayama:2008wy}. Figure \ref{fig1} also shows the GWB in the presence of a prolonged reheating phase, where we assumed a large reheating period lasting till the onset of big-bang nucleosynthesis (BBN).
	
	This analysis shows that the predicted GWB will not be observable with the current sensitivity of GW observatories, nor by third-generation interferometers. Also, the long reheating scenario is ruled out because it violates the BBN upper bound. This constraint, however, can be bypassed assuming fewer e-folds of contraction. Considering a contraction period of $\cN=55$ instead of 60 e-folds, which is the minimum conservatively assumed to isotropize the universe, is enough to avoid violations of the BBN bound \cite{Ben-Dayan:2019gll}.
	
	
	\section{String-gas cosmology with Atick--Witten conjecture}\label{sec3}
	
	One of the characteristic features of string theory is the existence of the Hagedorn phase at temperatures close to the string scale $M_{\rm s}$, where the energy is not dominated by the massless modes but rather by the most massive string states, leading to a pressureless fluid \cite{Deo:1988jj,Deo:1989bv,Brandenberger:1988aj,Tseytlin:1991xk}. In fact, a canonical description of the thermal phase indicates a limiting Hagedorn temperature \cite{Deo:1988jj,Deo:1989bv}. However, it was also argued that the limiting temperature only corresponds to the emergence of a thermal tachyonic mode, making the description of the system in terms of fundamental string excitations invalid \cite{Sathiapalan:1986db,Kogan:1987jd}. Atick and Witten~\cite{Atick:1988si} conjectured that, at temperatures larger than the Hagedorn temperature $T_{\rm H}$, the free energy grows much more slowly than in conventional field theories. Calculations in closed bosonic string theory \cite{Atick:1988si} indicated that $T_{\rm H}$ is not an upper bound but a point of first-order phase transition to a thermal ensemble with fewer degrees of freedom than expected from the zero-temperature string spectrum or from standard point-like particle field theories. This leads to the Atick--Witten scaling $\sim T^2$ of the free energy and, consequently, to a $p=\rho$ equation of state for the matter content of the early Universe. Such a background also arises from other, somewhat independent considerations in a setting obeying the holographic principle, where the $p=\rho$ equation of state saturates the entropy bound \cite{Banks:2001px,Banks:2003ta}.
	
	According to \cite{Atick:1988si}, the partition function only grows as $T^2$, hence the authors modeled the pressure with quadratic, linear and logarithmic terms in temperature. However, for convenience and transparency we may as well model the pressure in the form \cite{Biswas:2014kva}
	\begin{equation}\label{Talfa}
		p(T)= M_s^4\left[\left(\frac{T}{M_s}\right)^2+c_1\left(\frac{T}{M_s}\right)^{1+\alpha}\right]\,,\qquad |\alpha|\ll 1\,.    
	\end{equation}
	where $\a$ is a real constant.
	
	The possibility of thermal fluctuations being the origin of small inhomogeneities and anisotropies in the CMB was already proposed by Peebles~\cite{Peebles:1994xt}. The fluid fluctuations may arise naturally from two different sources. There might be fluctuations in the energy density and the associated temperature. And, even if there is a unique temperature in a given volume, there are fluctuations in energy within this volume due to the very statistical nature of thermal physics. This could also potentially seed primordial fluctuations; see, for instance, \cite{Magueijo:2002pg,Magueijo:2007wf,Wu:2009tt,Cai:2009rd} and references therein.
	
	The statistical fluctuations in the energy inside a given volume $V=L^3$ is given by
	\begin{eqnarray}
		\langle\Delta E\rangle_L^2&\coloneqq & \langle E^2\rangle-\langle E\rangle^2=\frac{\partial^2 \ln Z}{\partial \beta^2}=T^2 C_L\nonumber\\
		\Longrightarrow\quad \langle\delta \rho^2\rangle_L&=& \frac{T^2 C_V}{L^6}= \frac{T^2}{L^{3}}\frac{ \p \rho}{\p T}\,,
	\end{eqnarray}
	where $C_L$ is the heat capacity of the thermal system for the volume $L^3$.  These are classical random fluctuations that exist in any finite-temperature system as long as the fluid is in local thermal equilibrium. Hence, as such there is no need for any seed quantum fluctuations here.\footnote{Even in the vacuum dominated case, the initial fluctuations can be seeded classically to mimic the Bunch--Davies vacuum~\cite{Ashoorioon:2012kh}.} Therefore, the power spectrum for the seed fluctuations could then be sourced by these sub-Hubble wavelengths before the bounce \cite{Biswas:2014kva}. Once the modes become super-Hubble, thermal correlations over the relevant physical wavelengths can no longer be maintained, then the coupled  metric and matter fluctuations can evolve according to the usual general-relativistic hydrodynamical differential equations. 
	
	A precise understanding of how these statistical fluctuations get encoded in the curvature perturbation $\zeta$ at the Hubble crossing, was achieved in \cite{Biswas:2013lna} for a general extensive thermodynamic fluid whose pressure $p$ can be an arbitrary function of the temperature $T$. The derived curvature power spectrum reads
	\begin{eqnarray}
		{\cal P}_{\zeta}=k^3 \zeta_k^2  =  8 \sqrt{3\pi^3}A^2(T_k)\frac{T_k^2\rho_k'}{\Mpl^3\sqrt{\rho_k}}\,,\label{spectrum}
	\end{eqnarray}
	where a prime denotes differentiation with respect to $T$ and
	\be
	A(T)=\frac{3(1+w)\Omega + 2(3+R)}{6(1+w)\Omega}\,,\qquad R=-\frac{3}{2}\left[1+\frac{( 1+w)\rho( 2\rho'+T\rho'')}{T{\rho'}^2} \right],
	\ee
	where $\Om=\rho/(3\Mpl^2H^2)$. The subscript $k$ (which we are going to subsequently drop) refers to the fact that all these quantities have to be evaluated at horizon crossing, $H_k=k/a$. Note, that all the above functions of temperature can be calculated if we know $p(T)$, as the energy density is related rather straightforwardly to pressure:
	\begin{equation}
		\rho(T)=T\frac{d p(T)}{d T}-p(T)\,.
	\end{equation}
	
	The primordial tensor and scalar spectra as functions of the temperature $T$ are \cite{Biswas:2014kva}
	\ba
	\cP_{\rm t}(k) &=& \frac{2}{\pi^2}\left[\frac{H(k)}{\Mpl}\right]^2,\label{taw}\\
	\cP_{\rm s}(k) &=& \frac{\sqrt{3\pi^3} c_1^2}{4}\left(\frac{M_{\rm s}}{\Mpl}\right)^3\left[\frac{T(k)}{M_{\rm s}}\right]^{2\a},\label{Ps}
	\ea
	where $M_{\rm s}$ is the string mass scale and $c_1$ is the constant coefficient in the $O(T^{1+\a})$  term in \Eq{Talfa}. By tuning the parameters $c_1$ and $M_{\rm s}$ in \Eq{Ps}, one can easily make the tensor-to-scalar ratio
	\be
	r=\frac{8}{\pi^3\sqrt{3\pi} c_1^2}\left(\frac{H}{T}\right)^2\frac{\Mpl}{M_{\rm s}}\left(\frac{T}{M_{\rm s}}\right)^{2(1-\a)},
	\ee
	as large as the upper bound $r=0.036$ at $k_*=0.05\,{\rm Mpc}^{-1}$. This, however, turns out to be ruled out by observations, as we shall see presently.
	
\begin{figure}[ht]
		\centering
		\includegraphics[width=12cm]{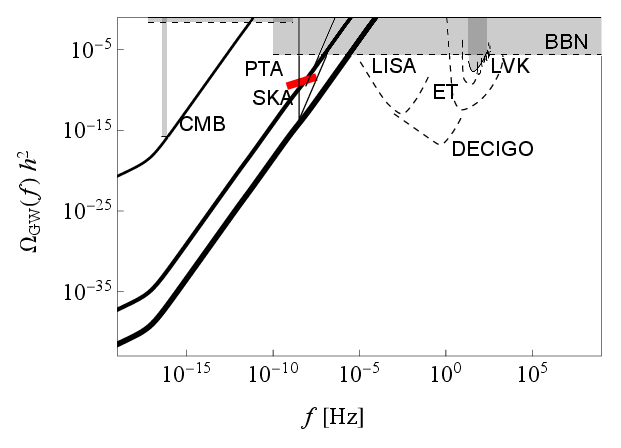}\\
		\includegraphics[width=12cm]{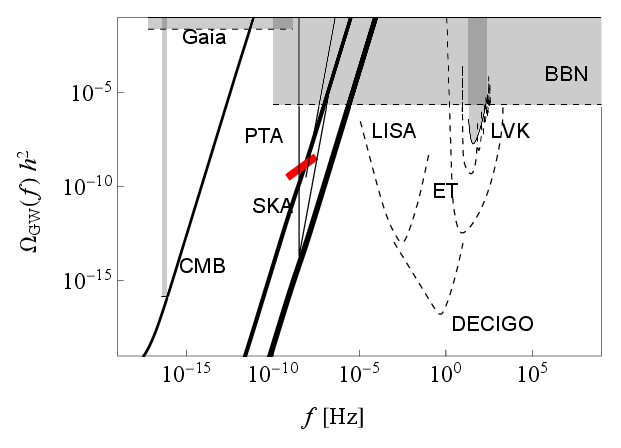}
		\caption{GWB for the bouncing model of string-gas cosmology with Atick--Witten conjecture with $r=0.036,10^{-18},5\times 10^{-22}$ (increasing thickness) at the CMB pivot scale (where the upper bound is given thickness to improve visibility in the plot), compared with the sensitivity curves (dashed) of LVK, SKA, LISA, ET and DECIGO. Also shown are the BBN exclusion region and, as a thick red segment, the signal detected by IPTA. The bottom plot captures the high-amplitude part of the GWB.\label{fig2}}
	\end{figure}

	At high temperature, the string gas behaves like a stiff fluid $p\simeq\rho$ and the gravitational setting is Einstein's gravity, so that the standard Friedmann equations hold and
	\be\label{grreg}
	H^2\propto \rho\propto T^2\propto a^{-6}\,.
	\ee
	Therefore, the energy scale of the model is $T_{\rm H}\sim M_{\rm s}\lesssim T\sim |H|$. From the horizon crossing relation $k=a|H|$, we have $k\propto a^{-2}$ and
	\be
	\cP_{\rm t}\propto a^{-6}\,,\qquad\cP_{\rm s}\propto a^{-6\a}\,,\qquad\frac{\rmd}{\rmd\ln k}= -\frac12 \frac{\rmd}{\rmd\ln a}\,,
	\ee
	so that the tensor and scalar spectral indices \Eq{nsnt} are constant and given by
	\be\label{ntnsaw}
	n_{\rm t} =3\,,\qquad n_{\rm s}-1 =3\a\,.
	\ee
	The tensor spectrum \Eq{taw} can thus be written as
	\be\label{Ptkei}
	\cP_{\rm t}(k)=r(k_*)\,\cP_{\rm s}(k_*)\left(\frac{k}{k_*}\right)^{n_{\rm t}},
	\ee
	which we plugged into \Eq{Omgw} to get the spectral shape shown in Fig.\ \ref{fig2}.
	
	As the reader can see, the signal predicted by this model is too strong and is excluded by all present constraints. The problem is the high tensor index in \Eq{ntnsaw}, two orders of magnitude higher than the minimum tilt required for detection in DECIGO, ET and LISA ($n_{\rm t}\gtrsim 0.06\!-\!0.34$) \cite{Calcagni:2023vxg}. At $r=0.036$, the curve hits the Gaia bound around $\Om_\textsc{gw}(f=5\times 10^{-12}\,{\rm Hz})\approx  7\times 10^{-12}$ \cite{Jaraba:2023djs}. The tensor-to-scalar ratio must be reduced down to $r\sim 10^{-18}$ to cross the signal detected by PTA, with which it is compatible.\footnote{Compare the reconstructed posterior distribution for $n_\textsc{gw}\simeq n_{\rm t}$ in Fig.\ 5 of \cite{NANOGrav:2023hvm}, where the tilt $n_\textsc{gw}$ of the GWB $\Omega_{\textsc{gw}}(f)\sim f^{n_\textsc{gw}}$ coincides with the primordial tensor tilt $n_{\rm t}$ in this case. The string-gas tilt $n_{\rm t}=3$ falls well within the likelihood region for a wide range of reheating temperatures.} To shift past the IPTA-SKA curve, it must be $r<5\times 10^{-22}$, at which point the parameter space of the model becomes too compressed to yield a realistic scenario. In the meanwhile, the BBN bound $\Om_\textsc{gw}< 5\times 10^{-6}$ truncates the top of all these signals before they can reach the cut-off frequency, which is much larger than the range depicted in the plot \cite{Calcagni:2020tvw}.
	
	As we warned in the introduction, this model is phenomenological and there are several points which need further theoretical study. First, a matter bounce takes place thanks to T-duality in string theory, which relates short to large distances $R\leftrightarrow 1/R$. However, just around the bounce event, one assumes Einstein gravity for simplicity. There are examples of quantum gravity where the inception of the Einstein regime is extremely close to the Planck scale \cite{Modesto:2022asj} but, in general, in the deep ultraviolet regime above the Hagedorn temperature gravity might be described by some modified action. However, to extract such dynamics is a difficult and yet unsolved problem. This issue, taken on board also in previous literature on the subject \cite{Brandenberger:2011et,Brandenberger:2012um}, will obviously need further study.
	
	Second, from the scale hierarchy reported below \Eq{grreg} is clear that the modes exiting the horizon during the contraction phase before the bounce have very high momenta. For these modes to stretched to cosmological scales, a mechanism for extending these wave-lengths from the Planck scale (or some order of magnitude away from it) to present-day low-multipole scales must be implemented in the model.
	In \cite{Biswas:2014kva}, some tentative solutions to this problem were advanced. One is to have a phase of inflation with mild slow roll stretching the wave-lengths to cosmological scales. This phase would be insufficient \emph{per se} to sustain a long-enough early-universe acceleration with the observed CMB spectrum. However, it would constitute a watering down of the original idea of having an actual alternative to inflation. Another stretching mechanism could involve a cyclic phase of growth \cite{Biswas:2014kva}. None of this has been explored in detail and will be critical to have under control the reformulation of the primordial tensor power spectrum \Eq{taw} as the power law \Eq{Ptkei}.
	
	
	\section{Pre-big-bang cosmology}\label{sec4}
	
	A possible example of primordial tensor perturbations peaked at high frequencies, with a strongly blue-tilted low-frequency regime of spectrum, is provided by the so-called Pre-Big-Bang (PBB) scenario \cite{Gasperini:1992em,Gasperini:2002bn}, based on the scale-factor duality of the string cosmology equations \cite{3}. Such symmetry is a peculiar property of string theory, and is a crucial ingredient not only to fix the slope of the primordial spectrum \cite{4,5,6} but also, as confirmed by recent results \cite{7,8}, to implement a smooth transition from the initial growing curvature (PBB) regime to the standard regime of decreasing curvature evolution.  
	According to this model, the evolution of our Universe, characterized by decelerated expansion (at intermediate times), decreasing temperature and curvature, weak gravitational coupling, should be preceded in time by an almost specularly symmetric phase of accelerated expansion, increasing curvature, increasing  density and temperature and growing coupling. Such a dual counterpart of the present one describes a ``pre-big-bang'' evolution from a flat, cold, empty initial state with negligible interactions to a final high-curvature, high-energy, explosive bounce, marking the transition to the more standard cosmological regime (see, e.g., \cite{Gasperini:2021mat} for a recent non-technical introduction, and \cite{16} for a more detailed and complete discussion).
	
	Here we recall the derivation of the associated GWB in a self-contained way. We shall introduce the spectral energy density of the relic GW radiation present today inside our cosmic horizon, and produced by a simple model of PBB scenario which satisfies all present observational constraints and depends on four constant parameters (see also \cite{Gasperini:2016gre}). Two of these parameters control the inflationary growth of the scale factor and of the string coupling in the high-energy regime preceding the bounce; the other ones control the beginning and the end of the axion-dominated phase occurring after the curvature bounce. We recall that the presence of a dust-like phase dominated by the oscillations of the Kalb--Ramond axion is in general needed in the PBB scenario to obtain (via the curvaton mechanism) a flat spectrum of scalar metric perturbations \cite{10,11}.
	
	At present, the first two parameters are largely arbitrary, while the other two may vary in a rather small range of values. We stress, however, that the dependence of the amplitude of the GWB on the full set of the above four parameters is given and discussed, for the first time, in this paper. We have neglected a possible further parameter, the effective propagation speed of tensor perturbations during the high curvature string phase, as it  seems to have small effects on the energy density of the GW spectrum.\footnote{On the contrary, the effective sound speed of scalar curvature perturbations and of axion perturbations is important for the production of primordial black holes, as discussed, e.g., in \cite{12} for the PBB scenario.}

	Let us now compute the spectral energy density in critical units of the GWB, \Eqq{41} with 
	$k_0 \leq k \leq k_1$, in such a way that
	all modes $k$ of eq.~(\ref{41}) satisfy the condition $k \tau_0 \gaq 1$ (i.e., they are all inside our present Hubble horizon, $k \gaq k_0= \tau_0^{-1}$). The highest mode $k_1= \tau_1^{-1}$ is the maximum amplified frequency crossing the horizon just at the end of the phase of PBB inflation. Higher frequency modes ($k \gg k_1$) can, in principle, be included into Eq.~(\ref{41}), but their amplitude is exponentially suppressed \cite{13,14,15} and their contribution to $\Om_{\textsc{gw}}$ is negligible. For the explicit computation of $\Om_{\textsc{gw}}$, two comments must be made. 
	
	The first is that, even if we are working in general with a higher-dimensional spacetime manifold (an unavoidable choice in the string theory context), we are mainly interested in the tensor perturbations of the four-dimensional metric, $\da g_{\mu\nu}= h_{\mu\nu}$, assuming that the extra spatial dimensions are today compactified with frozen dynamics. However, this does not mean that we are neglecting the possible effects of the higher-dimensional geometry during its initial, non-trivial evolution: indeed, all such higher-dimensional contributions will be included into the canonical equation which controls the dynamics of $h_{\mu\nu}$ \cite{Gasperini:2002bn,Gasperini:2016gre,16}.
	
	The second point is that, as usual, we are interested in the contributions to $\Om_{\textsc{gw}}$ arising from the cosmological amplification of the quantum vacuum fluctuations of the metric tensor. This implies that we can describe the amplification of tensor perturbations as a quantum (or semi-classical) field-theory process of production of pairs of gravitons from the initial vacuum state (see, e.g., \cite{16}) and we can write, for each mode $k$, the differential energy density of the amplified perturbations as follows:
	\beq
	\rmd \r_k(\tau_0) = 2 k\,\langle n_k (\tau_0) \rangle \,\frac{\rmd^3 \bm{k}}{(2 \pi)^3} =
	\frac{k^4}{\pi^2} \,\langle n_k (\tau_0) \rangle \,\rmd \ln k\,,
	\label{42}
	\eeq
	where $2$ is the number of polarization states and $\langle n_k (\tau_0) \rangle$ the number density of produced gravitons at the final epoch $\tau_0$. The last equality follows from assuming an isotropic final distribution. 
	
	To obtain $\langle n_k  \rangle$ and then $\Om_{\textsc{gw}}$, we need now to solve the evolution equation for the Fourier component of the (Mukhanov-Sasaki) canonical variable, $v_k(\tau)$, defined by putting in canonical form the quadratic action for the tensor fluctuation mode $h_k$ \cite{17}:
	\beq
	v_k'' + \left( k^2 - \frac{\xi''}{\xi} \right) v_k=0\,.
	\label{43}
	\eeq
	Here a prime denotes differentiation concerning the conformal time, $v_k= \xi h_k$ and $\xi(\tau)$ is the so-called ``pump field'' which controls, according to the above equation, the dynamics of the perturbation modes in the given background.
	
	In the model we are considering, the background may be approximated as a sequence of various cosmic phases, and in each of them the pump field $\xi$ is characterized by a simple power-law behaviour. In particular, in the initial PBB regime $- \infty < \tau \leq -\tau_1$, starting asymptotically from the string perturbative vacuum and approaching the curvature bounce at $\tau = - \tau_1$, we have two different phases with the following (canonically normalized) pump field behaviour \cite{Gasperini:2016gre}:
	\beq
	\xi \sim \frac{\Mp}{\sqrt 2} ( -\tau)^{1/2}, \qquad \tau < -\tau_s~ ;
	~~~~~~~~~~~~
	\xi \sim \frac{\Mp}{\sqrt 2} ( -\tau)^{\b-1}, \!\!\!\qquad -\tau_s < \tau < -\tau_1\,.  
	\label{44}
	\eeq
	The parameter $\b$ describes the high-energy growth of the dilaton and the dynamics of the internal dimensions, while the time scale $\tau_s$ is a free parameter which marks the transition from the low-energy initial phase to a possible late-time attractor, where spacetime curvature stays frozen at the value controlled by the fundamental string mass scale \cite{18}. In both phases, the evolution of the pump field takes into account not only, as usual, the inflationary growth of the scale factor but also the additional string-theory effects \cite{Gasperini:2002bn,16}, such as the dynamics of the extra dimensions and the growth of the string coupling controlled by the scalar dilaton field (see also \cite{18,19}). 
	
	In the subsequent post-bouncing regime $- \tau_1 < \tau \leq \tau_0$, the cosmic evolution is decelerated and we may have in principle three phases with the following pump-field behaviour \cite{Gasperini:2016gre}:
	\beq\label{45}
	\xi \sim \frac{\Mp}{\sqrt 2} \,\tau,~~ -\tau_1 < \tau < \tau_\sg; ~~~~
	\xi \sim \frac{\Mp}{\sqrt 2} \,\tau^2,  ~~ \tau_\sg < \tau < \tau_d;~~~~
	\xi \sim \frac{\Mp}{\sqrt 2} \,\tau,~~ \tau_d < \tau < \tau_{\rm eq}.  
	\eeq
	Here we are assuming that the extra-dimensional geometry and the string coupling (i.e., the dilaton) are frozen after the bounce, so that the pump field simply coincides with the scale factor. Here, again, we have two free parameters: the time scale $\tau_\sg$, marking the beginning of the dust-like phase dominated by the axion oscillations, and the time scale $\tau_d$, marking the epoch of axion decay associated with the conventional reheating (source of the CMB radiation that we are presently observing) and corresponding to the beginning of the standard post-big-bang evolution\footnote{There is also the final matter-dominated phase completing the cosmic evolution from the equality epoch $\tau_{\rm eq}$ down to the present epoch $\tau_0$. However, such a phase only affects the very low frequency modes $k < k_{\rm eq} = \tau_{\rm eq}^{-1}$, whose amplitude is so small (because of the strongly blue-tilted spectrum) to be fully negligible for this paper.}.
	
	As discussed in previous papers \cite{Gasperini:2016gre,10,11}, instead of $\tau_\sg$ and $\tau_d$ we can conveniently use as parameters the corresponding curvature scales $H_\sg \coloneqq H(\tau_\sg)$ and $H_d \coloneqq  H(\tau_d)$ which can be expressed in terms of the (unknown) mass $m$ of the Kalb--Ramond axion and of its initial amplitude $\sg_i$ after the bounce:
	\beq
	H_\sg  \simeq m \left(\frac{\sg_i}{\Mp}\right)^4, \qquad ~~~~~~~~~
	H_d \simeq m \left(\frac{m}{\Mp}\right)^2.
	\label{46}
	\eeq
	For a consistent model, the allowed values of the parameters $m$ and $\sg_i$ must satisfy the scale hierarchy $\Mp \gaq H_1 \gaq H_\sg \gaq H_d$. In addition, we have the obvious condition $H_d > H_N$, where $H_N \sim (1\, {\rm Mev})^2/\Mp$ is the nucleosynthesis scale of the standard cosmological scenario. 
	
	Given the full model of background evolution from the initial state at $\tau \ra -\infty$ down to the present epoch $\tau_0$, and given the power-law behaviour of $\xi$ in the various phases, we can now work in the so-called ``sudden approximation'' \cite{14} by imposing on the pump field to be continuous at the transitions scales, and solving, in each phase, the canonical equation (\ref{43}). We recall that, in general, for a pump field given by $\xi= (\Mp/\sqrt{2}) |\tau/\tau_1|^\a$, the exact solution for $h_k$ obtained from (\ref{43}) can be written in terms of the first- and second-kind Hankel functions, $H_\nu^{(1)}$, $H_\nu^{(2)}$, as
	\beq
	h_k(\tau) =\frac{v_k}{\xi}= \left(\frac{2 \tau_1}{\Mpl^2}\right)^{\frac12} \left | \frac{\tau}{\tau_1} \right|^\nu \left[ A_+(k) H_\nu^{(2)}(k\tau)+ A_-(k) H_\nu^{(1)}(k\tau)\right],
	\label{47}
	\eeq
	where $\nu = - \a +1/2$. The complete solution for $h_k(\tau)$, describing its evolution from $-\infty$ to $\tau_0$, is then obtained by solving the canonical equation in the various phases and matching $h_k$ and $h_k'$ at the  transition scales.
	
	In our model, in particular, we have four transitions (at $\tau_s$, $\tau_1$, $\tau_\sg$, $\tau_d$), which means five different phases of background evolution (see eqs.~(\ref{44}) and (\ref{45})), which implies five different solutions like (\ref{47}), and thus ten different coefficients $A_{\pm}(k)$ to be determined at the various epochs. The continuity of $h_k$ and $h_k'$ only gives eight conditions. The two remaining conditions are obtained by imposing on the canonical variable to initially describe a positive-frequency mode normalized to the quantum fluctuations of the Bunch--Davies vacuum, namely, by imposing $v_k = (1/\sqrt{2k}) \exp(-ik\tau)$ for $\tau \ra -\infty$: this implies (using the large-argument limit of the Hankel functions \cite{16}) $A_-=0$ and $A_+= \sqrt{\pi/4}$ for the solution describing perturbations in the initial regime $\tau \ra -\infty$. 
	
	With such a canonical normalization, the sought value of the number density $\langle n_k (\tau_0) \rangle$ of produced gravitons is then automatically obtained from the coefficient $A_-(k)$ of the first-kind Hankel function  describing the perturbation mode $h_k(\tau)$ in the final regime $\tau \ra +\infty$ (actually, for our purpose, in the limit $\tau \ra \tau_0$). More precisely, one finds (see, e.g., \cite{16})
	\beq
	\langle n_k (\tau_0) \rangle = \frac{4}{\pi} \left| A_-(k)\right|_{\tau =\tau_0}.
	\label{48}
	\eeq
	
	By performing the above computation and varying $k$ in the allowed frequency range we find that there are, in principle, four different branches of the energy density spectrum (\ref{41}), (\ref{42}), depending on the epochs of horizon crossing of the various modes. Noting that the axion-dominated phase is expected to occur soon after the bounce, in order to have a short duration with respect to the preceding high-curvature string phase (namely, $\tau_d/\tau_\sg \ll \tau_1/\tau_s$) we may consistently assume that all modes re-entering the horizon before, during, or soon after the axion phase are amplified modes leaving the horizon during the high-curvature string phase. This means, in other words, that we can work with the following hierarchy of wave-number scales: $k_1 \geq k_\sg \geq k_d > k_s$, where $k_i\coloneqq \tau_i^{-1}$ is the limiting frequency of a mode crossing the horizon at the transition epoch $\tau_i$.
	
	To obtain the explicit (parameter-dependent) form of the GWB, it is  useful to first  compute the frequency ratios of the four scales $k_i$. We note that in our model there are two phases of radiation-dominated evolution (i.e., $a \sim \tau \sim H^{-1/2}$) for $ -\tau_1 <\tau < \tau_\sg$ and for $\tau > \tau_d$, one phase of matter-dominated evolution (i.e., $a \sim \tau^2 \sim H^{-2/3}$) for $\tau_\sg <\tau < \tau_d$, and one phase where the string frame scale factor undergoes a de Sitter-like evolution (i.e., $a \sim |\tau|^{-1}$) for $-\tau_s <\tau < -\tau_1$. By defining the convenient parameters
	\be\label{zs}
	z_s \coloneqq \frac{\tau_s}{\tau_1}=\frac{k_1}{k_s}, ~~~~~~~~~~~~~
	z_\sg \coloneqq \frac{\tau_\sg}{\tau_1}=\frac{k_1}{k_\sg}, ~~~~~~~~~~~~~
	z_d \coloneqq \frac{\tau_d}{\tau_1}=\frac{k_1}{k_d},
	\ee
	controlling the time extension of the pre-bouncing high curvature regime and of the two post-bouncing, non-standard phases, we find\footnote{Conventions: we denote with $\om$ the usual (time-dependent) proper frequency evolving in time like the inverse scale factor, $\om(t) \equiv k/a(t)$. Hence, the proper frequency crossing the horizon at the given time $t_1$, $\om_1 = H_1 = H(t_1)$, and evaluated at a general time $t$, is given by $\om_1(t)= H(t_1) a(t_1)/a(t) \equiv H_1 a_1/a$. The time-dependent scale factor obviously disappears in the ratios of two frequency scales so that, for instance, $\om_1/\om_\sg= H_1 a_1/H_\sg a_\sg = k_1/k_\sg$.
	} (using eq.~(\ref{46}))
	\bea
	&&
	\!\!\!\!\!\!\!\!
	z_\s=\frac{k_1}{k_\sg }
	= \frac{H_1 a_1}{H_\sg a_\sg} = 
	\left(\frac{H_1}{H_\sg}\right)^{\frac12}
	\simeq\left(\frac{H_1}{\Mp}\right)^{\frac12}
	\left(\frac{m}{\Mp}\right)^{-\frac12}
	\left(\frac{\sg_i}{\Mp}\right)^{-2}, 
	\nonumber \\ &&
	\!\!\!\!\!\!\!\!
	\frac{z_d}{z_\s}=\frac{k_\sg}{k_d}= \frac{H_\sg a_\sg}{H_d a_d} = 
	\left(\frac{H_\sg}{H_d}\right)^{\frac13}\simeq 
	\left(\frac{m}{\Mp}\right)^{-\frac23}
	\left(\frac{\sg_i}{\Mp}\right)^{\frac43},
	\nonumber\\ &&
	\!\!\!\!\!\!\!\!
	\frac{z_s}{z_d}=\frac{k_d}{k_s}=
	\frac{k_d}{k_1}\frac{k_1}{k_s} =z_s \frac{H_d a_d}{H_1 a_1}=
	z_s \frac{H_d}{H_1} \frac{a_d}{a_\sg} \frac{a_\sg}{a_1} \simeq
	z_s \left(\frac{H_1}{\Mp}\right)^{-\frac12}
	\left(\frac{m}{\Mp}\right)^{\frac76}
	\left(\frac{\sg_i}{\Mp}\right)^{\frac23}. 
	\label{49}
	\eea
	By inverting the above relations we can also express $\s_i$ and $m$ in terms of $H_1$ and of the  parameters $z_s$, $z_\s$ and $z_d$ as follows:
	\be\label{zzz}
	\frac{\s_i}{\Mpl} \simeq \left(\frac{H_1}{\Mpl}\right)^{\frac16}z_d^{\frac14}z_\s^{-\frac{7}{12}}\,,~~~~~~~~\qquad \frac{m}{\Mpl} \simeq \left(\frac{H_1}{\Mpl}\right)^{\frac13}z_d^{-1}z_\s^{\frac13}.
	\ee
	
	Let us now give an example of full computation of the spectral distribution (\ref{41}) for the highest frequency branch of the spectrum ($k_\sg < k < k_1$) and for a generic epoch $\tau$ (late enough, however, to have all such modes inside the horizon, $k \tau \gg 1$). To this purpose, let us express the energy density of the perturbations in terms of their (time-dependent) proper frequency $\om$ scaling in time like the inverse scale factor, $\om(\tau) = k/a(\tau)$. From eqs.~(\ref{41}) and (\ref{42}), we obtain
	\beq
	\Om_{\textsc{gw}} (k,\tau) = 
	\frac{\om^4}{\pi^2 \r_{\rm crit}(\tau)} \,\langle n_\om (\tau) \rangle\,, ~~~~~~~~\qquad
	k_\sg < k < k_1.
	\label{410}
	\eeq
	The modes we are considering are amplified  by the pump field (\ref{44}) of the high-curvature string phase, and the canonical equations (\ref{47}), (\ref{48}) then give \cite{16}
	\beq
	\langle n_\om (\tau) \rangle \simeq \left(\frac{\om}{\om_1} \right)^{-1 - |3- 2 \b|}
	= \left(\frac{k}{k_1}\right)^{-1 - |3- 2 \b|}.
	\label{411}
	\eeq
	Also, it is convenient to express the critical  density $\r_{\rm crit}$ in terms of the critical fraction of radiation energy density, $\Om_{\rm r} = \r_r/ \r_{\rm crit}$, so that, referring to radiation produced at the axion-decay scale, we have
	\beq
	\r_{\rm crit}(\tau) = \frac{\r_r (\tau)}{\Om_{\rm r} (\tau)} = \frac{3 \Mp^2 H_d}{\Om_{\rm r} (\tau)}
	\left(\frac{a_d}{a}\right)^4.
	\label{412}
	\eeq
	Finally, for each mode of proper frequency $\om(\tau)$ we can express its $\om^4$ contribution to eq.~(\ref{410}) as
	\beq \om^4= \left(\frac{k}{a}\right)^4 = \left(\frac{k}{k_1}\right)^4 \left(\frac{H_1 a_1}{a}\right)^4.
	\label{413}
	\eeq
	Combining all these results and using eq.~(\ref{46}) we then find, for the considered band of frequency,
	\beq
	\Om_{\textsc{gw}}(k, \tau)= \frac{\Om_{\rm r} (\tau)}{\Om_{{\rm r}0}} \,\Om_{\rm PBB}\left(\frac{k}{k_1}\right)^{3- |3-2 \b|},
	~~~~~~~~~ k_\sg < k < k_1, ~~~~ \tau >\tau_d,
	\label{413a}
	\eeq
	where $\Om_{{\rm r}0}\equiv \Om_{\rm r}(\tau_0) \approx 4.15\times 10^{-5}h^{-2}$ is the present critical fraction of radiation energy density (including neutrinos) and we have defined the constant (parameter dependent) dimensionless amplitude
	\be
	\Om_{\rm PBB}\coloneqq \Om_{{\rm r}0}\left(\frac{H_1}{\Mp}\right)^2 \left(\frac{m \Mp}{\sg_i^2}\right)^{4/3}=\Om_{{\rm r}0}\left(\frac{H_1}{\Mp}\right)^2 \left(\frac{z_\s}{z_d}\right)^2.
	\ee
	For simplicity we have absorbed all numerical factors of order of unity into the unknown scale $H_1$. Obviously, the  result (\ref{413a}) is also valid if applied  in particular to the present epoch $\tau=\tau_0$, with $\Om_{\rm r} (\tau)= \Om_{{\rm r}0}$. 
	
	By following the same procedure for the other (lower frequency) branches of the spectrum, and turning to the more conventional frequency variable\footnote{Note that the initial normalisation of the spectrum to the  quantum fluctuations of the vacuum, leading to the results (\ref{48}), (\ref{410}), is performed as usual in units $\hbar=1$. Hence, $\om =2 \pi f$.}
	$f=k/(2\pi)$, we obtain that the full GWB can be written synthetically as
	\be
	\!\!\!\!\!\!\!\!\!\!\!\!\!\!\!
	\Om_{\textsc{gw}}(f) = \left\{
	\begin{matrix}
		\hspace{-.1cm}\Om_{\rm PBB}\left(\dfrac{f}{f_1}\right)^{3- |3-2 \b|},\qquad\qquad\qquad\qquad\qquad f_\sg \laq f \laq f_1 \\ \\
		\Om_{\textsc{gw}}(f_1)\left(\dfrac{f_\sg}{f_1}\right)^{3- |3-2 \b|}
		\left(\dfrac{f}{f_\sg}\right)^{1- |3-2 \b|},\qquad\quad  f_d \laq f \laq f_\sg \\ \\
		\Om_{\textsc{gw}}(f_\sg)\left(\dfrac{f_d}{f_\sg}\right)^{1- |3-2 \b|}
		\left(\dfrac{f}{f_d}\right)^{3- |3-2 \b|},\qquad\,\, f_s \laq f \laq f_d\\ \\
		\Om_{\textsc{gw}}(f_d)\left(\dfrac{f_s}{f_d}\right)^{3- |3-2 \b|}
		\left(\dfrac{f}{f_s}\right)^{3},\qquad \qquad \qquad\quad\,\,\,\,\, f \laq f_s
	\end{matrix}\right.
	\label{piecepbb}
	\ee
	where the three ratios of frequency scales $f_\sg/f_1$, $f_d/f_\sg$, $f_s/f_d$ can be expressed in terms of the parameters $z_s$, $z_\sg$, $z_d$ 
	(or $m$, $\sg_i$, $H_1$)
	according to \Eqq{49}. 
	We may note, for a better explanation of the above spectrum, that the higher frequency regime with interval $f \,\ep \,[f_\sg, f_1]$ concerns modes which exit the horizon in the string phase and re-enter in an early radiation-dominated phase, while modes with $f \,\ep \,[f_d, f_\sg]$ exit in the string phase but re-enter in a matter-dominated phase: hence the transfer function is different and yields the additional $f^{-2}$ spectral factor for $\b\approx 0$. Also, modes with $f \,\ep\, [f_s, f_d]$ exit the horizon in the same phase as before and re-enter in the standard big-bang radiation phase. Finally, the lowest frequency modes $f< f_s$ with the strongly blue spectrum $\Om \sim f^3$ exit the horizon in a low-energy phase which is stiff and contracting in the Einstein frame, with pump field $\xi(\tau) \sim (-\tau)^{1/2}$. Very simple  plots for the time evolution of the effective horizon, illustrating how modes belonging to  different spectral regimes may cross (in and out) the horizon, can be found, for instance, in \cite{16,16a}.

	It is also important to note that, for $f>f_1$, the spectrum is exponentially suppressed as $\Om_{\textsc{gw}}(f)=\Om_{\rm PBB}\exp{[-(f-f_1)/f_1}]$, so that a smooth interpolation of all the branches can be given by the following expression proven in Appendix \ref{appA}:
	\ba
	\Omega_{\textsc{gw}}(f)&=&\,\Om_{\rm PBB}\,f^{3}\left(f^{2}+f_{s}^{2}\right)^{-\frac{\left|3-2\beta\right|}{2}}\left(f^{2}+f_{d}^{2}\right)^{-1}\left(f^{2}+f_{\sigma}^{2}\right)\left(f^{2}+f_{1}^{2}\right)^{\frac{\left|3-2\beta\right|-3}{2}}\nn
	&&\times \exp{\left({-\frac{f}{f_1}+\arctan{\frac{f}{f_1}}}\right)}\,.
	\label{smoothpbb}
	\ea
	The smoothing of the piecewise profile \Eq{piecepbb} does not change the underlying scenario because the transition epochs from one phase to another are of very short, negligible duration compared with the time extension of such phases.
	
	There are now two important points to be stressed. The first is that the overall GWB, and in particular the peak amplitude, is controlled not only by the bouncing curvature scale $H_1$ but also by the parameters $m$ and $\sg_i$ and, thus, by the details of the post-bounce evolution. This implies, in particular, that the amplitude may result strongly suppressed with respect to the natural value fixed by the fundamental string mass scale $M_{\rm s}$, even for the highest frequency modes crossing the horizon at such scale. 
	
	The second point concerns the number of parameters controlling the shape of the spectrum. There are four time (or curvature) scales and one dimensionless number, the power $\b$ of the pump field in the string phase; see \Eq{44}. However, these five parameters must satisfy an important phenomenological condition. The PBB scenario we are considering, in fact, besides producing relic GW radiation must also produce a suitable large-scale background of scalar curvature perturbations with a nearly flat spectrum, in order to be compatible with CMB observations. 
	
	This is known to be possible via the curvaton mechanism \cite{20,21,22} triggered by the Kalb--Ramond axion \cite{10,11}, but this imposes  constraints on the previous spectral parameters \cite{Gasperini:2016gre}. In particular, the primordial scalar amplitude $\cP_{\rm s}(k_*)$ and spectral index $n_{\rm s}$ must be in agreement with the observational results reported below \Eq{nsnt}. By imposing such conditions on the scalar perturbations produced by the PBB model we are considering, whose spectrum is controlled by the same set of parameters as the GWB of \Eq{piecepbb} or its smooth version \Eq{smoothpbb}, we can then eliminate one of the previous five parameters and fix, for instance, the transition scale $H_1$ as a function of $z_s$, $\b$, $m$, $\sg_i$ and of the two observables $\cP_{\rm s}(k_*)$ and $n_{\rm s}$. This is done in Appendix \ref{appB1}, where we conclude that the parameter space of the model is
	\be\label{paspe}
	\{\b, z_s, z_d, z_\sg\}\,.
	\ee
	
	Without assuming any prior on these parameters, one can plot \Eq{smoothpbb} and find some general trends. For instance, the larger $\b$, the smaller the amplitude, so that, in practice, only values of $\b$ near zero generate a detectable signal. Also, as the parameter $z_d$ increases, the spectral shape is squeezed and the peak becomes narrower but neither its frequency nor its amplitude vary appreciably. On the other hand, the parameter $z_\s$ changes the shape but not much the frequency peak or the amplitude, while the parameter $z_s$ which affects both the range and the amplitude of the intermediate plateau or slope.
	\begin{figure}[h!]
		\centering
		\includegraphics[width=12cm]{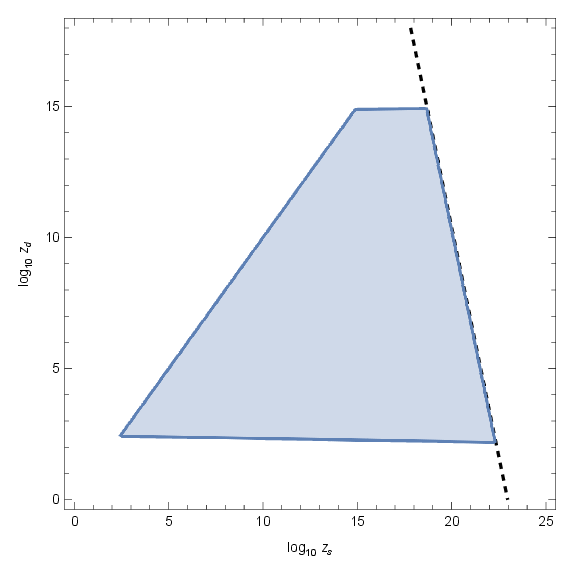}
		\caption{The shaded area shows the allowed region in the $(\log_{10}z_s,\log_{10}z_d)$ plane for the limiting case $\b=0$ and $\sg_i=\Mp$, which maximizes the peak amplitude of the primordial GWB. The maximal allowed intensity is reached for parameter values lying on the dashed straight line marking the right border of the region, corresponding to $\log_{10}(H_1/\Mp) \approx -3.29$, as explained in Appendix \ref{appB3}.}
		\label{fig3}
	\end{figure}
	
	However, the four parameters \Eq{paspe} do obey a non-trivial set of theoretical priors determined in Appendix \ref{appB2}. Within this space, in Appendix \ref{appB3} we circumscribe the region in the parameter space where the peak amplitude of the GWB is maximized. This happens for $\b=0$, $\sg_i=\Mp$, $z_\s$ determined by \Eqq{430} and the values of $z_s$ and $z_d$ shown in Fig.\ \ref{fig3}. 
	
	Any given set of parameters $z_s, z_d, z_\sg$ satisfying \Eqq{430} together with all other constraints, and implementing the additional limiting condition $\log_{10}(H_1/\Mpl) \approx -3.29$, produces a GWB with a peak of maximum intensity $\Omega_{\textsc{gw}}^{\rm max} \sim 10^{-11}\!-\!10^{-10}$ (\Eqq{432}). For phenomenological reasons, however, we are interested not only in the maximal amplitude but also in the maximal extension in frequency (in particular, towards small frequencies) of the allowed spectral region. This last property can be easily obtained by choosing, among all possible combinations of parameters producing the maximal amplitude,  
	the combination selecting the maximal allowed value of $z_s$ (i.e., $z_s \approx 10^{22.3}$), together with the corresponding minimal value of $z_d$ (i.e., $z_d \approx 10^{2.19}$), together with the associated  minimal value of $z_\sg$ (i.e., $z_\sg \approx 1$, according to \Eqq{430}).
	\begin{figure}[h!]
		\centering
		\includegraphics[width=12cm]{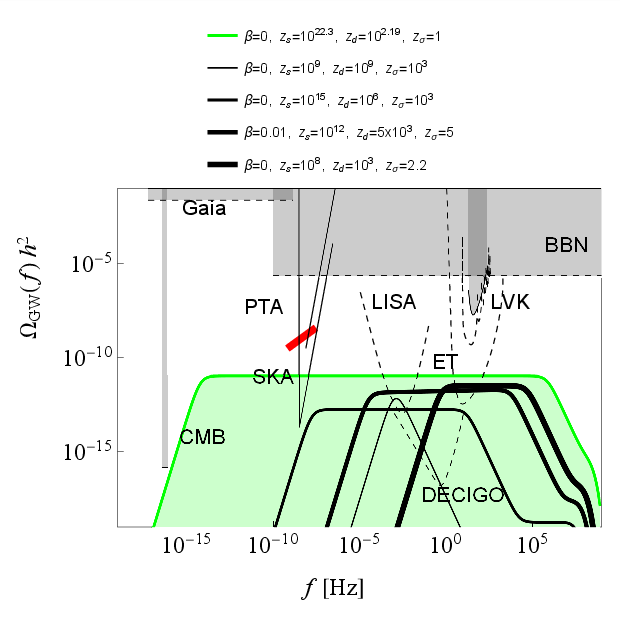}
		\caption{The shaded area is the maximal allowed region for the GWB \Eq{smoothpbb} extended in the low-frequency range, with all its parameters satisfying the self-consistency constraints (\ref{424})--(\ref{428}). We have also plotted four possible spectra corresponding to different sets of parameters giving rise to different kinematic details of the phases of earlier, non-standard cosmic evolution, preceding and following the curvature bounce.}
		\label{fig4}
	\end{figure}
	
	The GWB \Eq{smoothpbb} with the above numerical values of the parameters (satisfying $\sg_i=\Mp$) and with $\b=0$ provides the border of the shaded area in Fig.\ \ref{fig4} within which the primordial GWB of pre-big-bang cosmology with maximal peak amplitude falls. Such region respects known constraints obtained from the observations of millisecond pulsars \cite{pulsar}, which imply $\Omega_{\textsc{gw}} \laq 10^{-8}$ at a frequency scale $f \sim 10^{-8}$ Hz. It is also consistent with large-scale CMB constraints on the tensor spectrum. In the frequency range of IPTA, the maximal GWB amplitude is smaller than the signal detected \cite{NANOGrav:2023gor,EPTA:2023fyk,Reardon:2023gzh,Xu:2023wog,InternationalPulsarTimingArray:2023mzf} and is therefore consistent with those observations.
	
	For an illustrative purpose, in Fig.\ \ref{fig4} we have also plotted a few examples of GWBs produced by different sets of parameters satisfying all required constraints. The plotted spectra maximize neither the peak amplitude nor the extension in frequency but they are well inside the border of the allowed region and they can still produce a detectable signal within the LISA and the ET frequency range.
	

	\section{Discussion}\label{sec5}
	
	In this paper, we have continued the study by \cite{Calcagni:2020tvw} of the tensor-produced primordial GWB in bouncing scenarios embedded in, or inspired by, semi-classical models, quantum gravity and string theory. A grand summary of our past and present results is the following.
	\begin{itemize}
		\item Cosmological models of string and quantum gravity such as 
		inflation in flux compactification (see \cite{CQC,Baumann:2014nda} for reviews),
		nonlocal Starobinsky inflation \cite{Koshelev:2016xqb,Koshelev:2017tvv}, Wheeler--DeWitt quantum cosmology \cite{Kiefer:2011cc,Bini:2013fea,Kamenshchik:2014kpa,Kamenshchik:2015gua,Kamenshchik:2016mwj,Brizuela:2016gnz} and various incarnations of loop quantum cosmology \cite{Bojowald:2011iq,Agullo:2015tca,deBlas:2016puz,Gomar:2017yww,Li:2019qzr} generate a red-tilted primordial tensor spectrum ($n_{\rm t}<0$) and the resulting GWB is not enhanced with respect to the standard inflationary one. It is thus unobservable by present and planned interferometers \cite{Calcagni:2020tvw}.
		\item Scenarios such as Brandenberger--Ho non-commutative inflation \cite{Brandenberger:2002nq,Calcagni:2013lya},
		multi-fractional inflation \cite{Calcagni:2016ofu} and the conformal early universe in nonlocal quantum gravity \cite{Calcagni:2022tuz} also generate a blue tensor spectrum but the primordial GWB amplitude can only reach the DECIGO window \cite{Calcagni:2020tvw,Calcagni:2022tuz}. These models may allow for a strong signal at CMB scales already within observability reach but this is a fixed prediction only for the conformal early-universe scenario \cite{Calcagni:2022tuz}.
		\item The GWB of the S-brane ekpyrotic universe \cite{Brandenberger:2020tcr,Brandenberger:2020eyf,Brandenberger:2020wha} has a very low amplitude at CMB scales and it can reach the DECIGO window only if the tensor spectral index is running \cite{Calcagni:2020tvw}. In some extreme but not very realistic cases, it can even reach the ET window. The ekpyrotic scenario with Galileons and a $U(1)$ gauge field studied here \cite{Cai:2012va,Ben-Dayan:2016iks,Ben-Dayan:2018ksd,Artymowski:2020pci,Ben-Dayan:2023rlj} predicts a sourced tensor amplitude which is observable only on CMB frequency range, while the vacuum GWB is unobservable in all other upcoming GWB observations. It is likely that the same conclusions apply also to the new expyrotic scenario  \cite{Lehners:2007ac,Buchbinder:2007ad,Buchbinder:2007at,Buchbinder:2007tw,Qiu:2013eoa,Li:2013hga,Fertig:2013kwa,Ijjas:2014fja}, since the primordial tensor spectrum is the same \Eq{ekpyTT} and its steep slope upwards would result either in a very low tensor-to-scalar ratio or to a violation of the BBN bound, or to both.
		\item The GWB of string-gas cosmology is highly sensitive to the underlying assumptions. If the signal is produced below the Hagedorn temperature \cite{Brandenberger:2006xi,Brandenberger:2011et,Brandenberger:2012um,Brandenberger:2014faa,Bernardo:2020nol,Bernardo:2020bpa}, then the tensor spectrum is mildly blue-tilted and its associated GWB can reach at most the DECIGO sensitivity window assuming small or no running of the spectral indices \cite{Calcagni:2020tvw}. If, as explored here, it is produced above the Hagedorn temperature \cite{Biswas:2014kva}, then it becomes too high and is observationally excluded. A possible way out of this would be to revise the assumptions underlying the model, in particular, to be already in Einstein's gravity regime \Eq{grreg} when the primordial spectra were produced. Relaxing this condition and allowing generation of the spectra before recovering Einstein's gravity might yield different results. In section \ref{sec3}, we also noted some theoretical holes in the model that should be filled in order to understand whether our conclusions on its phenomenology are robust.
		\item The only model so far striking a balance between observability in the high-frequency range and observational consistency is the pre-big-bang scenario \cite{Gasperini:1992em,Gasperini:2002bn,
			Gasperini:2007vw,Gasperini:2016gre}. We found that its primordial GWB has a convex shape with an intermediate flat plateau. The parameter space of the model is tightly constrained theoretically but it still gives enough phenomenological freedom, to the point that this GWB can comfortably fall within the sensitivity window of both LISA and ET. When the parameter $\b$ is close to zero, in this range of frequencies the GWB of this model is a single power law
		\be\label{spl}
		\Omega_{\textsc{gw}}(f)\sim f^{n_\textsc{gw}}\,,\qquad n_\textsc{gw}=2+n_{\rm t}\simeq 3-|3-2\b|\approx 0\,.
		\ee
		When $z_d$ grows, the signal becomes a broken power law,
		\be
		\Om_{\textsc{gw}}(f) \sim \left\{
		\begin{matrix}
			\,\,f^{n_{\textsc{gw},1}},\qquad\quad f \ll \bar f \\
			f^{n_{\textsc{gw},2}},\qquad\quad  f \gg \bar f 
		\end{matrix}\right.,
		\label{piecepbb2}
		\ee
		where the slopes depend on where the transition scale $\bar f$ lies in the model's scales hierarchy and on whether the intermediate plateau is extended enough:
		\ba
		\hspace{-2.cm}&&\bar f= f_d\,:\hphantom{\simeq f_d}\quad n_{\textsc{gw},1}\simeq 3-|3-2\b|\approx 0\,,\quad\,\, n_{\textsc{gw},2}\simeq 1-|3-2\b|\approx -2\,,\nn
		\hspace{-2.cm}&&\bar f= f_s\,:\hphantom{\simeq f_d}\quad n_{\textsc{gw},1}\approx 3\,,\qquad\qquad\qquad\qquad n_{\textsc{gw},2}\simeq 3-|3-2\b|\approx 0\,,\label{bpl}\\
		\hspace{-2.cm}&&\bar f=f_s\simeq f_d\,:\quad\!\! n_{\textsc{gw},1}\approx 3\,,\qquad\qquad\qquad\qquad n_{\textsc{gw},2}\simeq 1-|3-2\b|\approx -2\,.\non
		\ea
	\end{itemize}
	Overall, among all these early-universe scenarios only the pre-big-bang model is of interest for LISA and ET. Both the single- and the broken-power-law shapes \Eq{spl} and \Eq{bpl} fit the two simplest templates commonly used for inflationary models \cite{Kuroyanagi:2018csn} and they could be submitted to the same type of analysis performed in \cite{peinf} as well as to a discussion on degeneracies and model selection when compared with alternative candidates with similar spectral shapes. 
	
	In general, the higher the signal-to-noise ratio, the smaller the error on the parameters of the spectral shape, as has been verified for a variety of cases \cite{Calcagni:2023vxg,Kuroyanagi:2018csn,peinf}. Thus, for the PBB model one would be able to identify the corner of the parameter space such that the tilt or tilts of the GWB (and, thus, $\beta$) would be determined with an accuracy of, say, 1\% by LISA and ET in their respective observability window. 
	
	In the case of the broken power law, the sharper the transition from one slope to the other with respect to the sensitivity curve the more difficult the determination of the parameters, due to the fact that the binning of the data introduces a certain level of coarse graining of short-range features \cite{Caprini:2019pxz}. We do not expect to have this problem for the PBB model, since the aforementioned transition is gentle enough (Fig.\ \ref{fig4}). These and other aspects of the PBB model (like, for instance, the possible contributions of shear and bulk viscosity arising at high-energies and affecting the high frequency regimes of the spectrum \cite{noi}) will deserve to be explored in a near future.
	
	
	\section*{Acknowledgments}
	
	G.C.\ is supported by grant PID2020-118159GB-C41 funded by the Spanish Ministry of Science, Innovation and Universities MCIN/AEI/10.13039/501100011033. He thanks S.\ Kuroyanagi and K.\ Schmitz for useful discussions. M.G.\ and E.P.\ are supported in part by INFN under the program TAsP: ``{\it Theoretical Astroparticle Physics}.'' E.P.\ is also partially supported by the research grant number 2022E2J4RK PANTHEON: ``{\it Perspectives in Astroparticle and Neutrino THEory with Old and New messengers }'' under the program PRIN 2022 funded by the Italian Ministero dell’Università e della Ricerca (MUR). I.B.-D., U.T.\ and A.V.\ are supported in part by the ``Program
	of Support of High Energy Physics'' Grant by Israeli Council for Higher
	Education; they also acknowledge the Ariel HPC Center at Ariel University for providing computing resources that have contributed to the research results reported in this paper.
	
	\appendix
	\section{Smooth interpolation of a broken power law}
	\label{appA}
	
	Suppose we have a piecewise continuous function $f(x):\mathbb{R}^{+}\rightarrow\mathbb{R}^{+}$ such that, in each interval, its behaviour is given by a power law $f(x)\sim x^{m}$. Then, we can define a sequence of exponents $m_{1},m_{2},\dots$
	and interval extrema $x_{1,2},x_{2,3},\dots$ such that $x_{i,i+1}<x_{i+1,i+2}$. The derivative of the logarithm of the function with respect to the logarithm of its argument is simply given by
	\begin{equation}
		\frac{d\ln f(x)}{d\ln x}=m_{1}+(m_{2}-m_{1})\Theta(\ln x-\ln x_{1,2})+(m_{3}-m_{2})\Theta(\ln x-\ln x_{2,3})+\dots,\label{eq:A1}
	\end{equation}
	\\
	where $\Theta$ is the Heaviside step function: $\Theta(y)=1$ for $y\geq0$
	and $\Theta(y)=0$ otherwise. Let us approximate the Heaviside function
	with a logistic function:
	\begin{equation}
		\Theta(y-y_{0})\simeq\frac{1}{1+e^{-l(y-y_{0})}},\label{eq:A2}
	\end{equation}
	\\
	where $l$ is a positive constant and the greater $l$ the better
	the approximation. Substituting the latter into (\ref{eq:A1})
	and defining a different $l$ for each $\Theta$, we have in compact
	form
	\begin{equation}
		\frac{d\ln f(x)}{d\ln x}=m_{1}+\sum_{i=1}^{N-1}\frac{m_{i+1}-m_{i}}{1+(\frac{x}{x_{i,i+1}})^{-l_{i+1}}}\,.\label{eq:A3}
	\end{equation}
	We can moreover rewrite the last equality with respect to the $x$-derivative as
	\begin{equation}
		\frac{d\ln f(x)}{dx}=\frac{m_{1}}{x}+\sum_{i=1}^{N-1}\frac{m_{i+1}-m_{i}}{x}\frac{1}{1+(\frac{x}{x_{i,i+1}})^{-l_{i+1}}}\,.\label{eq:A4}
	\end{equation}
	Integrating the last expression, we obtain
	\ben
	\ln f(x)=\ln A+\ln(x^{m_{1}})+\sum_{i=1}^{N-1}\ln(x^{l_{i+1}}+x_{i,i+1}^{l_{i+1}})^{(m_{i+1}-m_{i})/l_{i+1}},
	\een
	where $A$ is a constant, and finally 
	\ba
	f(x)&=& A\,x^{m_{1}}\prod_{i=1}^{N-1}\left(x^{l_{i+1}}+x_{i,i+1}^{l_{i+1}}\right)^{(m_{i+1}-m_{i})/l_{i+1}}\nn
	&=& A\,x^{m_{1}}\left(x^{l_{2}}+x_{1,2}^{l_{2}}\right)^{(m_{2}-m_{1})/l_{2}}\left(x^{l_{3}}+x_{2,3}^{l_{3}}\right)^{(m_{3}-m_{2})/l_{3}}\times\dots.\,
	\label{eq:A6}
	\ea
	
	If we wish to introduce an exponential cut-off for $x>x_{N,N+1}$ such
	that $f(x)\sim\exp[-(x-x_{N,N+1})/x_{N,N+1}]$, then we have to add
	to (\ref{eq:A1}) an additional contribution given by $(-x/x_{N,N+1}-m_{N+1})\Theta(\ln x-\ln x_{N,N+1})$. Let us denote $x_{N,N+1}\equiv x_{M}$ and
	$l_{N+1}\equiv l$. Following the same steps as before, one finds
	upon integration and exponentiation that
	\ba
	f(x)&=& A\,x^{m_{1}}\prod_{i=1}^{N-1}\left(x^{l_{i+1}}+x_{i,i+1}^{l_{i+1}}\right)^{(m_{i+1}-m_{i})/l_{i+1}}\left(x^{l}+x_{M}^{l}\right)^{-m_{N+1}/l}\nn
	&& \times\exp\left[-\frac{(x/x_{M})^{1+l}\,_{2}F_{1}(1,\,1+1/l,\,2+1/l,\,-(x/x_{M})^{l})}{1+l}\right],\label{eq:A7}
	\ea
	where $_{2}F_{1}(a,b,c,z)$ is the Gauss hypergeometric function.
	Translating into the general language we introduced before for the
	power spectrum of \eqref{piecepbb}, we have $m_{1}=3$, $m_{2}=3-\left|3-2\beta\right|$,
	$m_{3}=1-\left|3-2\beta\right|$ , $m_{4}=3-\left|3-2\beta\right|$,
	$x_{1,2}=f_{s}$, $x_{2,3}=f_{d}$ and $x_{3,4}=f_{\sigma}$ and $x_{M}=f_{1}$,
	so that (setting $l_{i}=l$ for all $i$)
	\ba
	\Omega_{\text{GW}}^{\text{smooth}}(f)&=&A\,f^{3}\left(f^{l}+f_{s}^{l}\right)^{-\frac{\left|3-2\beta\right|}{l}}\left(f^{l}+f_{d}^{l}\right)^{-\frac{2}{l}}\left(f^{l}+f_{\sigma}^{l}\right)^{\frac{2}{l}}\nn
	&&\times \left(f^{l}+f_{1}^{l}\right)^{\frac{\left|3-2\beta\right|-3}{l}}\mathcal{F}(f,f_{1},l)\,,\label{eq:A8}
	\ea
	where we defined the function $\mathcal{F}(f,f_1,l)$ as the exponential cutoff appearing in the last line of (\ref{eq:A7}). Setting $l=2$, we obtain (\ref{smoothpbb}):
	\ba
	\Omega_{\textsc{gw}}^{\text{smooth}}(f)&=&A\,f^{3}\left(f^{2}+f_{s}^{2}\right)^{-\frac{\left|3-2\beta\right|}{2}}\left(f^{2}+f_{d}^{2}\right)^{-1}\left(f^{2}+f_{\sigma}^{2}\right)\left(f^{2}+f_{1}^{2}\right)^{\frac{\left|3-2\beta\right|-3}{2}}\nn
	&&\times \exp{\left({-\frac{f}{f_1}+\arctan{\frac{f}{f_1}}}\right)}.\label{eq:A9}
	\ea
	The integration constant $A$ is fixed to match the asymptotic behaviour in the limit $f\rightarrow0$ with the explicit formula given in (\ref{piecepbb}),
	\begin{equation}
		\Omega_{\textsc{gw}}(f)\stackrel{f\rightarrow0}{\simeq}{\Omega}_{\text{PBB}}\,(f_{\sigma})^{2}(f_{d})^{-2}(f_{s})^{-\left|3-2\beta\right|}(f_{1})^{\left|3-2\beta\right|-3}f^{3},\label{eq:A10}
	\end{equation}
	while for the smooth interpolation \eqref{eq:A9}
	\begin{equation}
		\Omega_{\textsc{gw}}^{\text{smooth}}(f)\stackrel{f\rightarrow0}{\simeq}A\,(f_{\sigma})^{2}(f_{d})^{-2}(f_{s})^{-\left|3-2\beta\right|}(f_{1})^{\left|3-2\beta\right|-3}f^{3},\label{eq:A11}
	\end{equation}
	which yields $A={\Omega}_{\text{PBB}}$.

	
	\section{Parameter space of the PBB model}
	\label{appB}
	
	In this appendix, we show that the parameter space of the PBB model is $\{\b, z_s, z_d, z_\sg\}$ (section \ref{appB1}) and we discuss the theoretical priors on these parameters (section \ref{appB2}). We also determine the region in parameter space for which the promordial GWB amplitude is maximized (section \ref{appB3}). In view of a numerical implementation of these conditions and to emphasize orders of magnitude, it may be useful to work with base-10 logarithmic expressions.
	
	
	\subsection{Reducing the number of parameters}\label{appB1}
	
	First, we show that the transition scale $H_1$ is not independent and can be fixed by the other parameters,
	\be
	H_1=H_1(\b, z_s, z_d, z_\sg)\,.
	\ee
	By using previous results \cite{Gasperini:2016gre,12}, obtained under the  natural assumption that the pivot scale belongs to the low-frequency band of the scalar spectrum (i.e., $k_* < k_s$), the condition on $H_1$ following from the normalization of the scalar spectrum can be written as 
	\ba
	\left(\frac{H_1}{\Mp}\right)^{\frac{5-n_{\rm s}}{2}}&=& \frac{2 \pi^2}{\cT^2 (\sg_i)} \cP_{\rm s}(k_*)\, z_s^{1- n_{\rm s}-2 \b} \left[\left(\frac{H_*}{\Mp}\right)^{-\frac12} 
	\left(\frac{m \Mp}{\sg_i^2}\right)^{\frac13}\right]^{n_{\rm s}-1}\nn
	&=&\frac{2 \pi^2}{\cT^2 (\sg_i)} \cP_{\rm s}(k_*)\, z_s^{1- n_{\rm s}-2 \b} \left(\frac{H_*}{\Mp} 
	\frac{z_d}{z_\s}\right)^{-\frac{n_{\rm s}-1}{2}},
	\label{416}
	\ea
	where we have recast $m$ and $\s_i$ according to \Eq{zzz},
	$H_*$ is the curvature scale at the epoch in which the pivot mode $k_*$ re-enters the horizon and $\cT(\sg_i)$ is the transfer function connecting the amplitude of the primordial axion fluctuations to the final amplitude of the scalar curvature modes of metric perturbations. A numerical integration of the scalar perturbation equations gives the simple result \cite{11}
	\beq
	\cT(\sg_i) \simeq 0.13 \,\frac{\sg_i}{\Mp} + 0.25 \,\frac{\Mp}{\sg_i} -0.01\,,
	\label{417}
	\eeq
	where $\s_i$ can be expressed in terms of the $z$ parameters as in \Eq{zzz}. To obtain $H_*$, we can conveniently refer to the equilibrium scale by noting that $k_* \simeq 5 k_{\rm eq}$. This implies $H_*^{1/2} \simeq 5 H_{\rm eq}^{1/2}$. On the other hand, it is known that the Hubble parameter at radiation-matter equality is given by $H_{\rm eq} \simeq 1.6\times 10^{5} H_0\approx 9.5\times10^{-56}\Mpl$. We thus obtain
	\be\label{H*M}
	\left(\frac{H_*}{\Mp}\right)^{1/2} \approx 1.5\times 10^{-27}\,.
	\ee
	Now we can then express the normalization \Eq{416} in terms of the four parameters $\{\b, z_s, z_d, z_\sg\}$ (and of known experimental numbers) as follows:
	\ba
	\log_{10} \left(\frac{H_1}{\Mpl}\right) &=& \frac{2}{5 - n_{\rm s}} \left\{ \log_{10} \left[\frac{4.2 \pi^2}{\cT^2(\sg_i)}\right] -9 + (1-n_{\rm s})(\log_{10}  1.5 - 27)\right.\nn
	&&  + \left.(1-n_{\rm s} -2 \b) \log_{10}  z_s +\frac{n_{\rm s}-1}{2} \left(\log_{10}  z_\sg - \log_{10}  z_d \right) \right\},\label{422}
	\ea
	where we have used $\cP_{\rm s}(k_*)=2.1 \times 10^{-9}$. 
	It should be noted that $\cT^2(\sg_i)$ also contains $H_1$ through \Eq{zzz} but the solution for $H_1$ can always be numerically obtained, in general, for any given set of values of the four independent parameters. 
	
	Finally, the other important quantity appearing in the GWB (\ref{piecepbb}) is today's value of the highest amplified  frequency mode $f_1$, which is given by 
	\ba
	f_1&=&\frac{\om_1(\tau_0)}{2\pi}=\frac{H_1a_1}{2\pi a_0}=\frac{H_1}{2\pi}\frac{a_1}{a_\s}\frac{a_\s}{a_d}\frac{a_d}{a_{\rm eq}}\frac{a_{\rm eq}}{a_0}\simeq\frac{H_1}{2\pi}\left(\frac{H_\s}{H_1}\right)^\frac{1}{2}\left(\frac{H_d}{H_\s}\right)^\frac{2}{3}\left(\frac{H_{\rm eq}}{H_d}\right)^\frac{1}{2}\left(\frac{H_0}{H_{\rm eq}}\right)^\frac{2}{3}=\nn
	&=&\frac{H_1^\frac{1}{2}}{2\pi}\left(\frac{z_\s}{z_d}\right)^\frac{1}{2}\frac{H_0^\frac{2}{3}}{H_{\rm eq}^\frac{1}{6}}\simeq\frac{3.9\times 10^{11}}{2\pi}\left(\frac{H_1}{\Mpl}\right)^\frac{1}{2}\left(\frac{z_\s}{z_d}\right)^\frac{1}{2}\,{\rm Hz}\,.
	\ea
	
	
	\subsection{Theoretical priors}\label{appB2}
	
	Having thus determined that the parameter space is $\{\b, z_s, z_d, z_\sg\}$, let us now turn to the priors we can impose on it theoretically.
	
	A first condition concerns the  parameter $\b$ controlling the power-law behaviour of the primordial GW spectrum at high frequencies, which is constrained to be in the range
	\be
	0 \leq \b <3\,.
	\label{424}
	\ee
	The lower limit is due to the assumption of growing string coupling (needed to implement a smooth bouncing transition \cite{18,7,8}), while the upper limit has to be imposed to avoid background instabilities \cite{25}.
	
	We have then a number of constraints following from the (already mentioned) 
	hierarchy of the transition frequency scales, which must satisfy the conditions
	$f_1\gtrsim f_\s>f_d>f_s$. They imply 
	\be
	1\lesssim z_\s<z_d<z_s\,.
	\label{425}
	\ee
	In addition, for an efficient implementation of the curvaton mechanism based on the oscillations of the Kalb--Ramond axion, the axion background field must be oscillating when it becomes dominant (at the curvature scale $H_\sg$).  From the axion dynamical equations, one finds \cite{10,11} that the oscillating regime starts at the scale $H_m \simeq m$. This leads to the condition $H_m \geq H_\sg$, which implies $\sg_i \leq \Mp$ and which, by using Eq. (\ref{zzz}), can be written in logarithmic form as
	\beq
	\log_{10}\left(\frac{H_1}{\Mp}\right) +\frac{3}{2} \log_{10} z_d -\frac{7}{2} \log_{10} z_\sg \leq 0\,.
	\label{426}
	\eeq
	
	Also, to be consistent with the established results of the post-inflationary scenario, we may expect that the reheating produced by the axion decay at the scale $H_d$, and marking the beginning of the standard cosmological evolution, occurs before the BBN scale, $H_\textsc{bbn} \simeq (1\,{\rm MeV})^2/\Mp$. This implies $H_d > H_\textsc{bbn}$ from which, using eqs. (\ref{46}) and (\ref{zzz}), we have the constraint
	\beq
	\log_{10}\left(\frac{H_1}{\Mp}\right) - 3 \log_{10} z_d+ \log_{10} z_\sg > - 42 -\log_{10} 4\,,
	\label{427}
	\eeq
	where we have used $\Mp \approx 2 \times 10^{18}$ GeV.
	
	Finally, the conditions concerning the scalar perturbation spectrum must be imposed not only at the pivot frequency scale $k_*$ but also, in principle, to all frequency scales included into the multipole expansion of the CMB anisotropy, and constrained by observational data. This means, in other words, that also the highest frequency modes $k_\textsc{lss}$ presently constrained by large scale structure (LSS) observations must be below the lowest frequency branch of the axion perturbation spectrum \cite{Gasperini:2016gre,12}, and this implies 
	$k_\textsc{lss} < k_s$, where $k_\textsc{lss} \sim 3\, {\rm Mpc}^{-1}\approx 60\, k_*$, namely
	$H_\textsc{lss}^{1/2} \simeq 60\, H_*^{1/2}$. The condition $k_\textsc{lss}/k_s =(k_\textsc{lss}/k_1)z_s <1$ then leads to yet another constraint that can be written as follows. Since
	\ban
	\frac{k_1}{k_\textsc{lss}} &=& \frac{H_1 a_1}{H_\textsc{lss}a_\textsc{lss}}=\frac{H_1}{H_\textsc{lss}}\frac{a_1}{a_\s}\frac{a_\s}{a_d}\frac{a_d}{a_\textsc{lss}}\simeq\frac{H_1}{H_\textsc{lss}}\left(\frac{H_\s}{H_1}\right)^{\frac12}\left(\frac{H_d}{H_\s}\right)^{\frac23}\left(\frac{H_\textsc{lss}}{H_d}\right)^{\frac12}\\
	&=& \left(\frac{H_1}{\Mpl}\right)^{\frac12}\left(\frac{H_*}{\Mpl}\right)^{-\frac12}\left(\frac{H_*}{H_\textsc{lss}}\right)^{\frac12}\left(\frac{H_\s}{H_d}\right)^{-\frac16},
	\ean
	from \Eqqs{46} and (\ref{zzz}) we get $H_\s/H_d=(z_d/z_\s)^3$ and from \Eq{H*M} we obtain
	\beq
	\log_{10} z_s < 26 - \log_{10} 9 +\frac{1}{2} \log_{10}\left(\frac{H_1}{\Mp}\right)+\frac{1}{2} \left( \log_{10} z_\sg - \log_{10} z_d \right)\,.
	\label{428}
	\eeq
	
	
	\subsection{Maximizing the signal}\label{appB3}
	
	Given the condition (\ref{422}) on $H_1/\Mpl$, the amplitude and the frequency distribution of the GWB \Eq{piecepbb} or \Eq{smoothpbb} are fully determined by $\b, z_s, z_d, z_\sg$. These four parameters are not completely free, as they must satisfy a non-trivial set of self-consistency conditions (Appendix \ref{appB}). Taking into account these constraints on the parameters, we can determine the maximal allowed region for the PBB signal in the spectral plane $(\Om_\textsc{gw}, f )$. Let us first notice that, thanks to the condition (\ref{424}), the GWB (\ref{smoothpbb}) may be possibly decreasing only in the frequency branch $f_d \leq f \leq f_\sg$. The peak of the spectrum may thus be located either at $f_1$ or at $f_d$, with corresponding amplitudes
	\beq
	\Omega_{\textsc{gw}}(f_1)=  \Omega_{{\rm r}0}\left(\frac{H_1}{\Mp}\right)^2\left(\frac{z_\sg}{z_d}\right)^2, ~~~~~~~~~
	\Omega_{\textsc{gw}}(f_d)=  \Omega_{{\rm r}0}\left(\frac{H_1}{\Mp}\right)^2
	z_d^ {|3-2 \b|-3}.
	\label{429}
	\eeq
	In the first case, given the constraints (\ref{425}), the maximal amplitude can be reached for the limiting values $z_d \simeq z_\sg$, which imply however $H_d \simeq H_\sg$: hence, in that case, the axion starts decaying as soon as it becomes dominant, and there is not enough time for an efficient curvaton mechanism. Also, in that case, the maximal amplitude would correspond to a frequency range $f \sim f_1$, in general too high for the sensitivity of present detectors.
	
	In the second case with the peak at $f_d$, given again the constraints (\ref{424}) and (\ref{425}), the maximal amplitude can be obtained  either in the limit $z_d \simeq z_\sg \ra 1$ or in the limit $\b \ra 0$. Discarding the first possibility (for the same reasons as before), in order to find the allowed region for the GW signal of maximal intensity we will thus concentrate on the limiting case $\b=0$ which, as we will see, automatically leads to a peak located in frequency ranges possibly accessible to third-generation detectors. 
	
	It should be noted, in addition, that the limiting amplitude reached at $f_d$ for $\b=0$ is only controlled by the ratio $H_1/\Mp$, whose maximal allowed value is bounded by the constraints (\ref{426}): hence, the amplitude of the GWB approaches its allowed maximum in the limit in which the condition (\ref{426}) is saturated by the equality
	\beq
	\log_{10} z_\sg = \frac{2}{7} \log_{10} \left(\frac{H_1}{\Mpl}\right) + \frac{3}{7} \log_{10} z_d\,.
	\label{430}
	\eeq
	This result has two important consequences.
	
	First of all, by using \Eqq{zzz}, we can check that the above condition is equivalent to the condition $\sg_i=\Mp$, and this uniquely fixes the transfer function (\ref{417}) leading to the constant numerical value $\cT^2 \approx 0.137$. Second, by inserting into the above condition the general expression (\ref{422}) for $H_1$, and solving for the variable $\log_{10} z_\sg$, we can eliminate $z_\sg$ everywhere and confine our discussion of the maximum allowed spectrum to a two-dimensional parameter space spanned by the variables $\{\log_{10} z_s, \log_{10} z_d \}$, with $\b=0$, $z_\sg$ given by \Eqq{430} and $H_1$ given by
	\ba
	\log_{10} \left(\frac{H_1}{\Mpl}\right) &\simeq& \frac{14}{37 - 9 n_{\rm s}} \left[\log_{10} \left(\frac{4.2 \pi^2}{0.137}\right) -9 + (1-n_{\rm s})(\log_{10}  1.5 - 27)\right.\nn
	&&  + \left.(1-n_{\rm s} ) \log_{10}  z_s +\frac{2}{7} (1-n_{\rm s}) \log_{10}  z_d \right].\label{431}
	\ea
	
	We can now easily impose all constraints (\ref{425})--(\ref{428}) and evaluate, in such a context, both the allowed region of parameter space and the maximal allowed value of the peak amplitude. It turns out (see Fig.\ \ref{fig3}), that the maximal value of \Eqq{431} compatible with the given constraints corresponds to $\log_{10}(H_1/\Mpl) \approx -3.29$, so that the expected maximum intensity of the primordial GWB is given by
	\beq
	\Omega_{\textsc{gw}}^{\rm max} =  \Omega_{{\rm r}0}\left(\frac{H_1}{\Mp}\right)^2 \approx 10^{-10.6}\,.
	\label{432}
	\eeq
	The allowed values  of the parameters compatible with this maximal intensity (and with the imposed constraints) are in the range $18.7 \laq \log_{10} z_s \laq 22.3$ and  
	$2.19 \laq \log_{10} z_d \laq 14.9$, as shown in Fig.\ \ref{fig3}. The corresponding value of $z_\sg$ is given by \Eqq{430} and lies in the range  $0 \laq \log_{10} z_\sg \laq 5.5$. 
	

\end{document}